\documentclass[12pt]{article} 
\usepackage{amsfonts}
\usepackage{amsmath}
\usepackage{amssymb}
\usepackage{subeqnarray}
\usepackage{epsfig}
\usepackage{amscd}
\allowdisplaybreaks[4]
\begin{document} 

\title{The many levels pairing Hamiltonian \\
for two pairs}
\author{M.B. Barbaro$^1$, R. Cenni$^2$, A. Molinari$^1$
and M. R. Quaglia$^2$\\
${}^1$ Dipartimento di Fisica Teorica, Universit\`a di Torino
\\
and INFN, Sez. di Torino, Torino, Italy \\
${}^2$ Dipartimento di Fisica, Universit\`a di Genova\\
and INFN, Sez. di Genova, Genova, Italy\\}
\date{}

\maketitle
\begin{abstract}
We address the problem of two pairs of fermions living on an arbitrary
number of single particle levels of a potential well (mean field) and
interacting through a pairing force in the framework of the Richardson's 
equations.
The associated solutions are classified in terms of a number $v_l$,
which reduces to the seniority $v$ in the limit of a large 
pairing strength $G$ and yields the number of pairs not developing a
collective behaviour, their energy remaining finite in the $G\to\infty$
limit.
We express analytically, through the moments of the single particle
levels distribution, the collective mode energy and the two critical values
$G_{\rm cr}^{+}$ and $G_{\rm cr}^{-}$ of the coupling which can exist
on a single particle level with no pair degeneracy.
Notably $G_{\rm cr}^{+}$ and $G_{\rm cr}^{-}$, when the number of 
single particle levels goes to infinity, merge into the critical coupling
of a one pair system $G_{\rm cr}$ (when it exists), which is not envisioned
by the Richardson theory.
In correspondence of $G_{\rm cr}$ the system undergoes
a transition from a mean field to a pairing dominated regime.
We finally explore the behaviour of the excitation energies, wave functions
and pair transfer amplitudes versus $G$ finding out that the former, for 
$G>G_{\rm cr}^{-}$, come close to the BCS predictions, whereas the latter
display a divergence at $G_{\rm cr}$, signaling the onset of a
long range off-diagonal order in the system.
\end{abstract}

\section{Introduction}
\label{sec:1}

In a previous paper \cite{BaCeMoQu-02} we examined in detail the problem of the
fermionic pairing Hamiltonian $\hat H_{pair}$ in the simple situation of one 
pair of nucleons coupled to an angular momentum $J=0$ living in a set of 
$L$ single particle levels (s.p.l.), e.g. 
in a major shell of the nuclear mean field. 
This problem was solved long
ago for the case of one s.p.l. hosting $n$ pairs of fermions: yet it
still presents aspects deserving further investigation in the non-degenerate
case.

Actually this is the situation one faces in applying $\hat H_{pair}$ to
real systems like nuclei and metals, which in fact represent two extreme
situations of the non-degenerate case: in the former
a major shell is typically split into five or 
six s.p.l. of different angular momenta, in the latter
the number of non-degenerate levels entering into a band corresponds to
a significant fraction of the Avogadro number. Moreover
in a heavy nucleus the number of pairs living in a level may be as large as,
say, eight while in a metal is one.

Recently a renewed and widespread interest for $\hat H_{pair}$
in the non-degenerate frame
 has flourished in connection
with the issues 
of the physics of ultrasmall metallic grains, possibly 
superconducting~\cite{Sierra:1999rc},
and of Bose-Einstein condensation~\cite{Dukelsky:2001fe}.

When only one pair in $L$ s.p.l. is considered then the eigenstates of
the pairing hamiltonian fall into two classes: one collective and
$L-1$ trapped (in between the s.p.l.) states. 
In this connection it was found in \cite{BaCeMoQu-02} that:

\begin{enumerate}
\item the energy $E_{coll}$ of the collective mode 
is related to the statistical features of the s.p.l. distribution, in the sense
that only few moments
of the latter, beyond of course the strength of the pairing interaction,
are sufficient to accurately predict $E_{coll}$;
\item the eigenvalues $E_{tr}^\nu$ ($\nu=2,\cdots L$) of the states
trapped in between the s.p.l. and hence constrained in the range
$e_{\nu-1}<E_{tr}^\nu<e_\nu$, being the $e_i$ ($i=1,\cdots L$) the single
particle energies, belong to two different regimes:
in the weak coupling regime, for $\nu$ both small and large,
the $E_{tr}^\nu$ almost coincide with $e_\nu$
(the more so, the smaller the degeneracy of the unperturbed s.p.l. is),
whereas in the strong coupling regime, {\em provided the degeneracy
of the s.p.l. increases with $\nu$}, the $E_{tr}^\nu$, for small $\nu$,
almost coincide with $e_{\nu-1}$.
Notably the transition between the two situations
occurs more and more sharply, namely in correspondence of a
precise value of the coupling constant $G$, as $L$ increases.
\end{enumerate}

In the present paper we extend the previous analysis
considering two pairs of fermions living in many levels. 
First we shall see that in this case the eigenstates of the pairing Hamiltonian
are conveniently classified in terms of a
number $v_l$, which provides a measure of the degree of collectivity of the 
states. 
This number is directly linked to the number $N_G$, first introduced by
Gaudin in ref.~\cite{Gau95}, through the relation $v_l=2 N_G$ and
might be considered as a sort of ``like-seniority'' (hence the notation
$v_l$)~\footnote{Note however that, whereas the seniority $v$ counts the
fermions coupled to $J\neq 0$, the number $v_l$ refers to $J=0$ pairs.} 
since it reduces to the standard seniority $v$ for large $G$: as 
pointed out in \cite{Roman}, it has the significance of the number of pair 
energies (see below) which remain finite as $G$ goes to infinity.

Futhermore for the $n=2$ system we address the issues of:
\begin{enumerate}
\item expressing the energy $E_{coll}$ of the collective $v_l=0$ state
in terms of the statistical features of
the s.p.l. distribution, i.e. in terms of the moments of the latter;
\item exploring the existence and providing an analytic expression
of the critical values of $G$ (we already know from \cite{Ric-65} 
that at most two of them might exist on a s.p.l.) which signal the
transition of the system between two different regimes;
\item relating these values of $G_{\rm cr}$ to the one previously found in
the $n=1$ case;
\item studying the transition from the weak coupling
(where the mean field dominates) to the strong coupling regime
(where the pairing interaction dominates) by following the behaviour
with $G$ of the pair transfer matrix elements between one and two pairs states;
\item relating the exact and BCS solution in the quite extreme situation of
few pairs (in fact two) and few s.p.l. (in fact three).
\end{enumerate}

Anticipating our results, we will find that indeed,
under suitable conditions, two values of $G_{\rm cr}$ exist.
They turn out to coincide in the large $L$ limit and, notably, 
they coincide as well with the one found in ref.~\cite{BaCeMoQu-02}
for the $n=1$ case.
Their analytic expression can also be related, as for $E_{coll}$,
to the moments of the distribution
of the s.p.l. (in fact to the inverse of the latter) and
they again mark the transition between the two above mentioned regimes.

Indeed for $G>G_{\rm cr}$ the exact solutions of the Richardson equations
appear to come very
close to the Bogoliubov quasi-particle solutions of the reduced BCS and, 
concerning the one pair transfer matrix element, we display their divergence
in the proximity of $G_{\rm cr}$.

Finally we prove that the existence of $G_{\rm cr}$ relates to the 
trace of $\hat H_{pair}$.

\section{Classification of the states}
\label{sec:class}

As well-known the eigenvalues and the eigenvectors of the pairing
Hamiltonian
\begin{equation}
\hat H_{pair} = \sum_{\nu=1}^L e_\nu \sum_{m_\nu=-j_\nu}^{j_\nu}
\hat a^\dagger_{j_\nu m_\nu} \hat a_{j_\nu m_\nu}
- G \sum_{\mu,\nu=1}^L \hat A^\dagger_\mu \hat A_\nu~,
\end{equation}
$\hat A_\mu=\sum_{m_\mu=1/2}^{j_\mu} (-1)^{j_\mu-m_\mu}
\hat a_{j_\mu,- m_\mu} \hat a_{j_\mu m_\mu}$ being the $J=0$ pair 
distruction operator,
are found by solving the Richardson equations \cite{Ric-64}.
These, for $n=2$, reduce to the following system of two equations
in the unknown $E_1$ and $E_2$
(referred to as pair energies):
\begin{subeqnarray}
1-G\sum\limits_{\mu=1}^{L} \dfrac{\Omega_{\mu}}{2e_{\mu}-E_1} + 
\dfrac{2G}{E_2-E_1} &=&0 \\
1-G\sum\limits_{\mu=1}^{L} \dfrac{\Omega_{\mu}}{2e_{\mu}-E_2} + 
\dfrac{2G}{E_1-E_2} 
&=&0
\label{sistema}
\end{subeqnarray}
or, equivalently, by adding and subtracting the above,
  \begin{subeqnarray}
2-G\sum\limits_{\mu=1}^{L} \dfrac{\Omega_{\mu}}{2e_{\mu}-E_1} -
G\sum\limits_{\mu=1}^{L} \dfrac{\Omega_{\mu}}{2e_{\mu}-E_2}&=&0 \\
-\sum\limits_{\mu=1}^{L} \dfrac{\Omega_{\mu}}{2e_{\mu}-E_1}
+\sum\limits_{\mu=1}^{L} \dfrac{\Omega_{\mu}}{2e_{\mu}-E_2} + 
\dfrac{4}{E_2-E_1} 
&=&0\,.
\label{sistema2}
\end{subeqnarray}
In (\ref{sistema},\ref{sistema2}) the degeneracy of the single particle
energy (s.p.e.) $e_\mu$ 
is $\Omega_\mu$ and $G$ is the strength of the pairing force.
Since the Richardson equations deal with pairs of fermions (nucleons)
coupled to an angular momentum $J=0$, their eigenvalues, given
by $E=E_1+E_2$, are those of the  
zero-seniority states ($v=0$). Importantly, these eigenvalues
display different degrees of collectivity: hence they are
conveniently classified in terms of the latter.
For this purpose, as mentioned in the Introduction, we introduce a number 
$v_l$,  
that counts in a given state the number of particles prevented
to take part into the collectivity, not because they are blind to the
pairing interaction (indeed they are coupled to $J=0$), 
but because they remain trapped in between the s.p.l., 
even in the strong coupling regime.
Specifically we shall ascribe the value $v_l$= 0 to the fully collective state,
$v_l$=2 to a state set up with a trapped pair energy while the other
displays a collective behaviour and $v_l$=4 to the state with
two trapped pair energies.

This classification is equivalent to the one of ref.~\cite{Roman},
where the $n$ pair energies are split (for any $n$) into two classes:
those remaining finite as $G\to\infty$ and the others.

We shall explore this pattern of states in both the weak and the strong
coupling regimes, commencing with the former, which is the
simpler to deal with.

\section{The weak coupling domain}
\label{sec:weak}

In the weak coupling limit of course no collective mode develops, hence
$v_l$ has no significance.
Adopting a perturbative treatment we write the pair energies
$E_1$, $E_2$ as
\begin{equation}
E_i= 2 e_{\mu_i} + G x_i
\end{equation}
($G x_i$ being a perturbation). Hence
\begin{equation}
\sum_{\mu}\dfrac{\Omega_\mu}{2 e_\mu -E_i}= 
-\dfrac{\Omega_{\mu_i}}{G x_i} + \sum_{\mu \ne \mu_i} 
\dfrac{\Omega_\mu}{2(e_{\mu}- e_{\mu_i})- G x_i}
\end{equation}
and, expanding in $G$, the system \eqref{sistema} becomes 
\begin{subeqnarray}
1+ \dfrac{\Omega_{\mu_1}}{x_1}-G\sum\limits_{\mu\ne \mu_1} 
\dfrac{\Omega_{\mu}}{2(e_{\mu}-
e_{\mu_1})} + 
\dfrac{2G}{2(e_{\mu_2}-e_{\mu_1})} + O(G^2)&=&0 \\
1+ \dfrac{\Omega_{\mu_2}}{x_2}-G\sum\limits_{\mu\ne \mu_2} 
\dfrac{\Omega_{\mu}}{2(e_{\mu}-
e_{\mu_2})} + 
\dfrac{2G}{2(e_{\mu_1}-e_{\mu_2})} + O(G^2)
&=&0\,,
\label{sistema2bis}
\end{subeqnarray}
where the indices $\mu_1$ and $\mu_2$ select one 
configuration out of the unperturbed ones.
At the lowest order in $G$, if $\mu_1\ne \mu_2$, 
one has $x_i=-\Omega_{\mu_i}$ and the pair energies
$E_i= 2e_{\mu_i}-G\Omega_{\mu_i}$ are real. Thus the energy eigenvalue
$E=E_1+E_2$ becomes
\begin{equation}
E= 2(e_{\mu_1} + e_{\mu_2}) - G (\Omega_{\mu_1}+ \Omega_{\mu_2})~.
\label{weakdiverso}
\end{equation}
Instead, if $\mu_1=\mu_2$, entailing $\Omega_{\mu_1}>1$,
we have to solve the system
\begin{subeqnarray}
1+ \dfrac{\Omega_{\mu_1}}{x_1}-G\sum_{\mu\ne \mu_1} 
\dfrac{\Omega_{\mu}}{2(e_{\mu}-
e_{\mu_1})} + \dfrac{2}{x_2-x_1} + O(G^2)&=0 \\
1+ \dfrac{\Omega_{\mu_2}}{x_2}-G\sum_{\mu\ne \mu_2} 
\dfrac{\Omega_{\mu}}{2(e_{\mu}-
e_{\mu_2})} + \dfrac{2}{x_1-x_2} + O(G^2)&=0\,.
\label{sistema3}
\end{subeqnarray}
A generic solution reads
\begin{equation}
x_{1,2}  = -(\Omega_{\mu_1}-1)\pm i \sqrt{\Omega_{\mu_1}-1} \,,
\label{sistema3bis}
\end{equation}
showing that $E_1$ and $E_2$ are always complex conjugate.
Hence the energy becomes 
\begin{equation}
E= 4 e_{\mu_1}  - 2 G (\Omega_{\mu_1}-1)\,,
\label{weakidentico}
\end{equation}
which is of course real.

In comparing \eqref{weakdiverso} and \eqref{weakidentico}
with the lowest energy of one pair, which, in the weak coupling regime, 
is~\cite{BaCeMoQu-02}
\begin{equation}
E= 2 e_{\mu_1} -G \Omega_{\mu_1}\,,
\label{weakuno}
\end{equation}
one sees that, while \eqref{weakdiverso} corresponds to the sum of two 
contributions like \eqref{weakuno} (the two pairs ignore each other), 
in \eqref{weakidentico} the Pauli blocking effect appears.

\section{The strong coupling domain}
\label{sec:strong}

Here is where $v_l$ has a significance and we deal
with the states with $v_l=0$ 
and $2$ (the $v_l=4$ states are of minor physical interest).

\subsection{$v_l=0$}

In this Subsection we study the eigenvalue of $v_l=0$ in the strong coupling 
limit.
Clearly such a state arises from an unperturbed configuration with the two
pairs in the lowest s.p.l. (if $\Omega_1>1$) or in the two lowest s.p.l.
We already know that in the degenerate case 
the collective eigenvalue is
\begin{equation}
E= 4 \bar e -2 G (\Omega-1)\,,
\end{equation} 
$\bar e$ being the energy and  $\Omega$ the degeneracy of the level. 
The above yields the leading order contribution to the energy 
in the non-degenerate case and represents a good estimate when
the spreading of the s.p.e. levels is small with respect to $G \Omega $, 
as it is natural to expect. 

To show this, following \cite{BaCeMoQu-02},
we introduce the new variables 
\begin{equation}
x_i=\frac{E_i-2\bar e}{G\Omega}
\end{equation}
where now
\begin{equation}
\bar e = \frac{1}{\Omega} \sum_{\mu=1}^L \Omega_{\mu} e_{\mu}~, 
~~~~\mbox{and}~~~~\Omega = \sum_{\mu=1}^L \Omega_{\mu}~.
\end{equation}

We further define the variance 
\begin{eqnarray}
  \sigma &=&\sqrt{\frac{1}{\Omega}\sum_\nu\Omega_\nu(e_\nu-\bar e)^2}
\end{eqnarray}
and the skewness
\begin{eqnarray}
\gamma&=&\frac{1}{\sigma^3\Omega}\sum_\nu\Omega_\nu(e_\nu-\bar e)^3\
\end{eqnarray}
of the level distribution; moreover we introduce the expansion parameter
\begin{equation}
  \label{eq:1}
  \alpha=\frac{2\sigma}{G\Omega}\,.
\end{equation}
Then the system \eqref{sistema2} becomes
\begin{subeqnarray}
2+\dfrac{1}{x_1}\left(1+\dfrac{\alpha^2}{x_1^2} + 
\dfrac{\alpha^3}{x_1^3}\gamma +\cdots\right)+ \dfrac{1}{x_2}
\left(1+\dfrac{\alpha^2}{x_2^2} + 
\dfrac{\alpha^3}{x_2^3}\gamma +\cdots\right)&=&0\nonumber\\~\\
\dfrac{1}{x_1}\left(1+\dfrac{\alpha^2}{x_1^2} + 
\dfrac{\alpha^3}{x_1^3}\gamma +\cdots\right)
- \dfrac{1}{x_2}
\left(1+\dfrac{\alpha^2}{x_2^2} + 
\dfrac{\alpha^3}{x_2^3}\gamma +\cdots\right)
+\dfrac{4}{\Omega}\dfrac{1}{x_2-x_1}&=&0\,.\nonumber\\
\label{sistema3ter}
\end{subeqnarray}
At the leading order  the system \eqref{sistema3ter} is easily 
solved, yielding
\begin{equation}
x_{1,2}^{(0)}=-\frac{(\Omega -1)}{\Omega} \pm i \frac{\sqrt{\Omega-1}}{\Omega}
\ .
\end{equation}
Higher order terms can be obtained through a recursive linearisation procedure.
One gets
\begin{eqnarray}
x_{1,2}&=&-\frac{(\Omega -1)}{\Omega}-
\alpha^2\frac{(\Omega-2)}{(\Omega-1)} +\alpha^3 \gamma 
\frac{(\Omega-4)}{(\Omega-1)}\nonumber\\
&\pm& i \frac{\sqrt{\Omega-1}}{\Omega}
\mp i \frac{1}{2} \alpha^2 \frac{\Omega}{(\Omega-1)^{3/2}}
\pm i \alpha^3 \gamma \frac{\Omega}{(\Omega-1)^{3/2}}
+{\cal O}(\alpha^4)\,.
\end{eqnarray}

Note that the two pair energies are always complex conjugate.
So the system's total energy reads
\begin{equation}
\frac{E-4 \bar e}{G \Omega}= -2\frac{(\Omega -1)}{\Omega}-
2\alpha^2\frac{(\Omega-2)}{(\Omega-1)} +2\alpha^3 \gamma 
\frac{(\Omega-4)}{(\Omega-1)} +{\cal O}(\alpha^4)~.
\label{energyn2}
\end{equation}

Since the collective energy of one pair of 
nucleons living in $L$ levels in the strong coupling regime is
\cite{BaCeMoQu-02}
\begin{equation}
\frac{E-2\bar e}{G \Omega} = -1 -\alpha^2 + \gamma \alpha^3+{\cal O}(\alpha^4)
\,,
\label{energyn1}
\end{equation}
one sees that, when $\Omega\gg 1$, \eqref{energyn2} 
becomes just twice the value \eqref{energyn1} and, moreover, the imaginary
part of $x_1$ and $x_2$ goes to zero as $1/\sqrt{\Omega}$.
Thus, in this limit, the Pauli interaction between the two pairs
vanishes, as expected: the two pairs behave as two 
free quasi-bosons condensed in a level 
whose energy is given by (\ref{energyn1}).

In Table \ref{tab:1} we compare the result \eqref{energyn2} with the exact 
one assuming that the two pairs live in the first $L$ levels of
a harmonic oscillator well with frequency $\omega_0$.
The $v_l=0$ collective state  arises from an
unperturbed configuration with a pair in the lowest and a pair
in the next to the lowest s.p.l. (in units of the oscillator frequency
the energy of such a configuration is 8).
We see that the difference between the two results never exceeds 
$\sim$~15~\%, even for $\tilde G=G/\hbar \omega_0$ as low as 0.1.
Note also that the energy of the $v_l=0$ state 
scales with the size of the well (i.e. it does not 
depend upon $\omega_0$).

\begin{table}[ht]
\begin{center}
\begin{tabular}{||c|c|c|c|c|c||}
\hline\hline  
$ \tilde G$  & $\alpha$ &   $E^{(0)}$   & $E^{(2)}$ & $E^{(3)}$ & 
$E^{\rm{exact}}$\\
\hline
0.1 & 0.89 & 11.2 & 8.21 & 6.01 & 7.25\\
0.2 & 0.44 & 7.4  & 5.91 & 5.36 & 5.30\\
0.3 & 0.29 & 3.6  & 2.61 & 2.36 & 2.31\\
0.4 & 0.22 & -0.2 & -0.95 & -1.08 & -1.11\\
0.5 & 0.18 & -4. & -4.60 & -4.69 & -4.70\\
0.6 & 0.15 & -7.8 & -8.30 & -8.36 & -8.37\\
0.7 & 0.13 & -11.06 & -12.03 & -12.07 & -12.08\\
0.8 & 0.11 & -15.4 & -15.77 & -15.81 & -15.81\\
0.9 & 0.10 & -19.2 & -19.53 & -19.56 & -19.56\\
1.0 & 0.09 & -23.0 & -23.30 & -23.32 & -23.32\\
1.5 & 0.06 & -42. & -42.20 & -42.21 & -42.21\\
\hline\hline
\end{tabular}
\end{center}
\caption{Energies, in units of $\hbar \omega_0$,
of the state $v_l=0$ for different values
of $\tilde G=G/\hbar \omega_0$ 
at the order 0, 2 and 3 in the expansion parameter 
$\alpha$ compared with the exact ones. 
The $L=4$ s.p.e. levels 
and the associated degeneracies are those of a 3-dimensional 
harmonic oscillator,  $\omega_0$ being 
the harmonic oscillator constant. \label{tab:1}}
\end{table}

\subsection{$v_l=2$}

\label{sec:vl2}

In the absence of the coupling term (Pauli principle)
in the Richardson equations \eqref{sistema} the eigenvalues of the $v_l=2$ 
states could be simply obtained by adding the collective energy $E_1$
carried by one pair
and the trapped energy $E_2$ carried by the other pair.

This situation is recovered in the very strong coupling limit,
where all the s.p.e. become essentially equal to 
$\bar e$ and both $\bar e$ and $E_2$ are very small
with respect to $E_1$. Indeed
the first equation of the system \eqref{sistema} then becomes
\begin{equation}
\frac{1}{G} + \frac{\Omega}{E_1} -\frac{2}{E_1}=0
\end{equation}
yielding 
\begin{equation}
E_1= -G(\Omega-2)~,
\end{equation}
namely the energy of the state with two pairs and $v=2$ 
in the one level problem. 
This result, setting a correspondence between states with $v_l=2$ and $v=2$,
connects {\em seniority}
and {\em ``like-seniority''} (or the physics of a
`broken' and a `trapped' pair). 

In the non degenerate case, denoting with $E^{(1)}_1$ and 
$E^{(\nu)}_1$ the first and the $\nu$-th 
eigenvalues ($\nu\not=1$) of the one pair equation
\begin{equation}
1-G\sum\limits_{\mu=1}^{L} \dfrac{\Omega_{\mu}}{2e_{\mu}-E_1} =0\,,
\label{eq:44}
\end{equation}
the coupling (Pauli) term in \eqref{sistema} can be approximated 
for large $G$ as follows
\begin{equation}
  \label{eq:46}
  \frac{2G}{E_2-E_1}\simeq\frac{2G}{E^{(\nu)}_2-E^{(1)}_1}\sim
  \frac{2G}{2e_\nu-2\bar e+ G\Omega}\sim\frac{2}{\Omega}\,.
\end{equation}
Hence the equations of the system
\eqref{sistema} decouple and can accordingly be recast as 
\begin{subeqnarray}
  \label{eq:49}
  \frac{1}{G_{\rm eff}^{(1)}}-\sum_\mu\frac{\Omega_\mu}
  {2e_\mu-E_1}=0\\
  \frac{1}{G_{\rm eff}^{(2)}}-\sum_\mu\frac{\Omega_\mu}
  {2e_\mu-E_2}=0\,,
\end{subeqnarray}
where
\begin{subeqnarray}
  \label{eq:50a}
  \frac{1}{G_{\rm eff}^{(1)}}
\equiv\frac{1}{G}+\frac{2}{E^{(\nu)}_2-E^{(1)}_1}
\simeq\frac{1+\dfrac{2}{\Omega}}{G}
\\
\label{eq:50b}
  \frac{1}{G_{\rm eff}^{(2)}}
\equiv\frac{1}{G}-\frac{2}{E^{(\nu)}_2-E^{(1)}_1}
\simeq\frac{1-\dfrac{2}{\Omega}}{G}
\,.
\end{subeqnarray}

We thus see that in the strong coupling regime the Pauli principle 
just re-scales the coupling constant, differently, however, 
for the collective and the trapped states: indeed the
pairing interaction is quenched for the former 
and enhanced for the latter by the Pauli blocking.

We compare in Table \ref{tab:2} the exact results with the approximate ones
(eq.~\eqref{eq:49}) 
in the h.o. case for $L=3$, 
in the strong coupling regime ($\tilde G=5$).
In this example five states exist: of these only two have $v_l=2$,
namely those associated with the unperturbed configurations having
either one pair in the lowest s.p.l. and one in the highest or both pairs in
the second s.p.l.
In the Table we see that 
the estimated trapped energy $E_2$ almost coincide with the exact ones, 
while the results for the collective energy $E_1$ 
are satisfactory (the error being $\simeq 7\% $).
 
\begin{table}[ht]
\begin{center}
\begin{tabular}{||c|c|c|c||}
\hline\hline  
$E_1^{\rm{exact}}$ & $E_2^{\rm{exact}}$ & $E_1$ & $E_2$ \\
\hline
-33.325 & 3.307 & -35.711 & 3.309\\
-33.988 & 5.720 & -35.711 & 5.721\\
\hline\hline
\end{tabular}
\end{center}
\caption{Exact and approximate (eq.~\eqref{eq:49}) 
energies of the states with $v_l=2$ for 
$\tilde G=5$. 
The s.p.e. and degeneracies are those 
of a 3-dimensional harmonic oscillator. 
All energies are in units of $\hbar\omega_0$. 
The number of levels considered is $L=3$. \label{tab:2}}
\end{table}

In concluding this Section, we observe that our results 
in the strong coupling domain agree with the findings of
reference \cite{Yuz03}, where the problem of the pairing Hamiltonian
for small superconducting grains is studied. 
Indeed, although in \cite{Yuz03} the single particle levels
are non-degenerate, the conclusions of this paper are not affected by this 
assumption.
Hence the results of ref.~\cite{Yuz03} for two pairs can be recovered from 
ours by setting $\Omega_\nu=1$ in the formulae of this Section.
Likewise, our results can be directly derived from ref.~\cite{Yuz03} by setting
$n$=2 and by grouping the single particle levels into $L$ degenerate 
multiplets.

\section{The critical value of $G$}
\label{sec:crit}

In this Section we first search for the
critical values of the coupling constant $G$ and for their expression in terms
of the moments of the s.p.l. distribution for the $n=2$ system
(Subsection \ref{sec:angcr}).
Next, in Subsection \ref{sec:Gcr}, we discuss the physics
of $G_{\rm cr}$.
Finally, in Subsection \ref{sec:BCS}, we compare the Richardson solution
with the BCS one.

\subsection{The analytic expression of $G_{\rm cr}$}
\label{sec:angcr}

In the weak coupling regime, when
a state evolves from an unperturbed one with the
two pairs living in the same level, then the pair
energies $E_1$ and $E_2$ are always {\em complex conjugate}.
On the other hand when the state evolves from an unperturbed one 
having the two pairs living in two different s.p.l., then 
$E_1$ and $E_2$ are {\em real}. 

By contrast, in the strong coupling regime
the pair energies $E_1$ and $E_2$ 
of the $v_l=0$ state are always complex conjugate.
It is then clear that, if the degeneracy $\Omega_1$ of the lowest s.p.l.
is greater than one, then the pair energies
$E_1$ and $E_2$ of the $v_l=0$ state are complex conjugate in both the
weak and strong coupling regime and their behaviour with $G$
is smooth.

On the other hand if $\Omega_1=1$, since in the weak coupling limit 
the two pairs must live on different levels, $E_1$ and $E_2$ are necessarily 
real in a neighbourhood of the origin, but become complex in the strong 
coupling regime.
Thus a singularity in their behaviour as a function of
$G$ is bound to occur. 

In this second case it appears natural to surmise that the
singularity takes place when $E_1$ and $E_2$ coincide.
In fact in this case the Pauli term of the Richardson equations diverges,
and it must be compensated by a divergence in the sum entering into the 
system (\ref{sistema}): this can only happen if $E_1$ or $E_2$
coincides with an unperturbed eigenvalue.

To find out the analytic expression of the critical value of $G$
we start from a generic solution 
with two pair energies $E_1$ and $E_2$,
which evolve with $G$ till they coincide at an unperturbed energy 
$2e_\nu$ of a s.p.l. with degeneracy $\Omega_\nu$.
These pair energies must fulfill the $G$-independent 
eq.~(\ref{sistema2}b), that we cast in the form
\begin{equation}
(E_2-E_1)^2 \sum_{\mu=1}^L \dfrac{\Omega_\mu}{(2 e_\mu-E_1)(2 e_\mu -E_2)} +4=0
\label{equazione}
\end{equation}
and which allows to express $ E_2$ as a function of $ E_1$.
To render this relationship explicit we set
\begin{subeqnarray}
\label{eq:ab36} 
E_1 & =& 2 e_\nu +x\\
E_2 & =& 2 e_\nu +x\varphi_\nu(x)\,,
\end{subeqnarray} 
where $\varphi_\nu(x)$ is easily found to read, for vanishing $x$,
$$\varphi_\nu(0)=\frac{\Omega_\nu-2\pm 2\sqrt{1-\Omega_\nu}}{\Omega_\nu}\,,$$
a real quantity only when $\Omega_\nu=1$. 
This occurrence is crucial since, in order to compensate the divergence arising
from the sum appearing in (\ref{sistema}a), namely
$-G\Omega_\nu/(2e_\nu-E_1)=G\Omega_\nu/x$, the divergent Pauli term
$2G/(E_2-E_1)=2G/[x(\varphi_\nu(x)-1)]$ must be real.

Assuming then $\Omega_\nu=1$, the link between the pair 
energies is easily found by inserting \eqref{eq:ab36} into \eqref{equazione} 
and expanding in $x$.
To fourth order in $x$ one obtains
\begin{equation}
  \label{eq:z1}
  \begin{split}
    E_2&=2e_\nu-x\pm 2{\cal P}_{(2)\nu} x^2-4{\cal P}_{(2)\nu}^2 x^3\\
    &+\left(\pm 9{\cal P}_{(2)\nu}^3+2{\cal P}_{(3)\nu}^3
      \pm\frac{{\cal P}_{(4)\nu}^4}{{\cal P}_{(2)\nu}}\right)
    x^4+{\cal O}(x^5)~,
  \end{split}
\end{equation}
where $E_1$ is hidden in $x$ and the quantities
\begin{equation}
  \label{eq:13n}
  {\cal P}_{(k)\nu}=\left\{\sum_{\mu=1(\mu\neq\nu)}^L
\frac{\Omega_\mu}{(2 e_\mu -2 e_\nu)^k}  \right\}^{\frac{1}{k}}
\end{equation}
are the inverse moments of the level distribution.

To compute $G_{\rm cr}$ we recast one of the equations of the system 
\eqref{sistema} as follows
\begin{equation}
  1+\frac{G}{x}-G \sum_{\mu=1(\mu\neq\nu)}^L \frac{\Omega_\mu}{2e_\mu-2e_\nu} 
-\frac{G}{x}\mp {\cal P}_{(2)\nu} G
  =0\,,
\label{equaz}
\end{equation}
which is valid in the $x\to 0$ limit and transparently displays the
cancellation of the divergences.
Clearly the two solutions of \eqref{equaz} are
\begin{equation}
  \label{eq:29n}
  G_{\rm cr}^{(\nu) \pm}=\frac{1}{{\cal P}_{(1)\nu}\pm
    {\cal P}_{(2)\nu}}=\left[\sum_{\mu=1(\mu\neq\nu)}^L 
    \frac{\Omega_\mu}{2 e_\mu-2 e_\nu}\pm 
    \sqrt{\sum_{\mu=1(\mu\neq\nu)}^L
    \frac{\Omega_\mu}{(2 e_\mu-2 e_\nu)^2}}\,\right]^{-1}~,
\end{equation}
which actually correspond to {\it  two} critical values for $G$.
We thus recover the results found long ago by Richardson~\cite{Ric-65}
through a somewhat different route.

\begin{figure}[hp]
  \begin{center}
    \epsfig{file=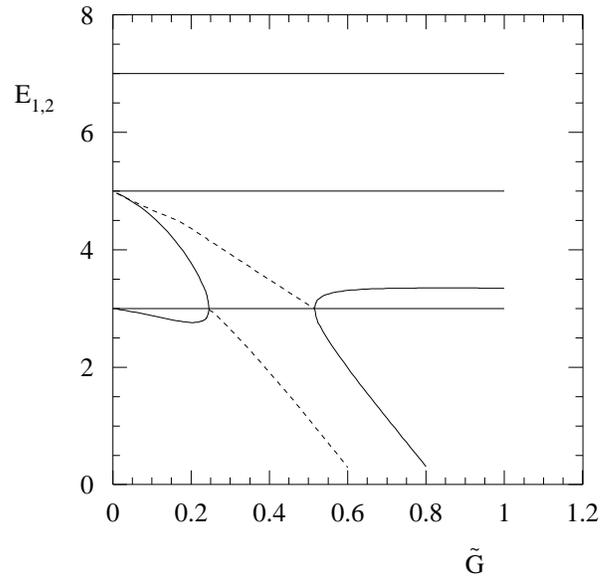,height=10cm,width=10cm}
    \caption{
Behaviour with $\tilde G$ of the pair energies of
the ground ($v_l=0$)
and first excited ($v_l=2$) states obtained as solutions of the Richardson
equations for a harmonic oscillator well for $L=3$:
dashed lines denote
the real part of the solutions (of course coincident) when the solutions
are complex and solid lines the separate real parts.
The pair energies are in units of $\hbar\omega_0$. As explained in the text
that the two critical
points refer to the two different states.
}
    \label{fig:2}
  \end{center}
\end{figure}

The situation is portrayed in
fig.~\ref{fig:2}, where the behaviour of the pair energies with 
$\tilde G$ is displayed 
for a harmonic oscillator well assuming $L=3$.
Clearly in this case the only s.p.l. with $\Omega=1$ is the lowest one, hence
the pair energies $E_1$ and $E_2$, real in the weak coupling limit, 
coincide at the critical point $\tilde G_{\rm cr}^{(1)+}$ 
(their common value being $2 e_1$)
and then become complex conjugate; the energy of the
associated state evolves in the $v_l=0$ collective mode. 
By contrast, for the $v_l=2$ state, arising from the configuration with two 
pairs living on the second level (which is allowed for 
the harmonic oscillator well), the two pair energies
$E_1$ and $E_2$ are complex conjugate in the weak coupling limit, coalesce
into the energy $2 e_1$ at the critical point $\tilde G_{\rm cr}^{(1)-}$ and then 
become real. 
One of the two solutions remains trapped above $2 e_1$, while the other 
evolves into a collective state: the sum of the two yields the energy
of the $v_l=2$ state.  
   
It is thus plain that 
for the occurrence of a critical value of $G$ 
a s.p.l. with $\Omega=1$ must exist.
Such s.p.l. is the lowest one in a harmonic oscillator, hence for this 
potential well both the ground and the first excited state of a  
$n=2$ system carry one critical value of the coupling constant, namely
$G_{\rm cr}^{(1)+}$ and $G_{\rm cr}^{(1)-}$, respectively.
In correspondence of these $G$ the pair energies take on the
value $E_1=E_2=2e_1$.
Thus for a $n=2$
system two (at most) critical points exist on a $\Omega=1$ s.p.l.

\begin{figure}[htp]
  \begin{center}
    \epsfig{file=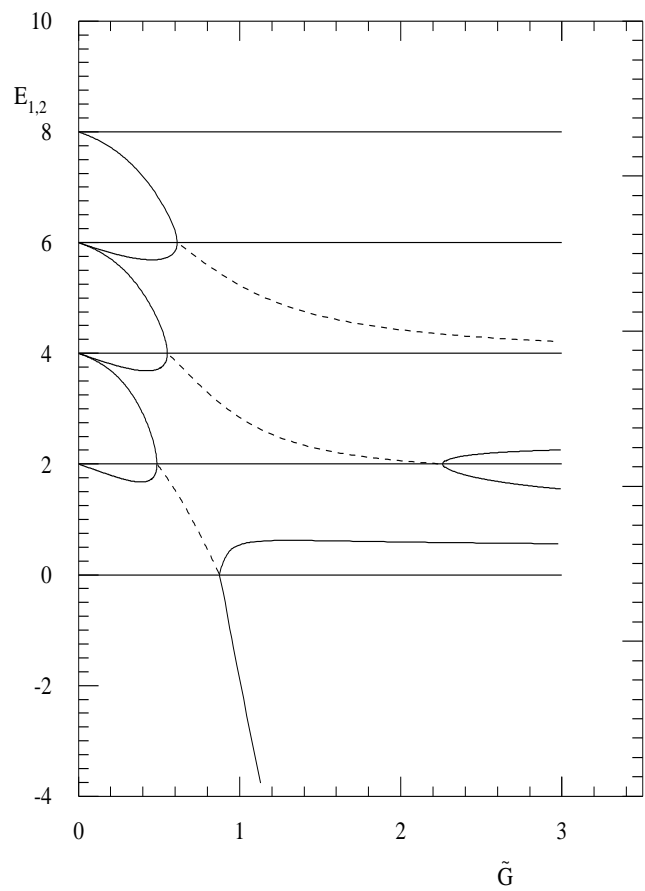,height=14cm,width=12cm}
    \caption{
Behaviour of the pair energies of three excited states
in the case of $\Omega_\mu=1$. The number of levels considered is 
$L=9$.}
    \label{fig:4}
  \end{center}
\end{figure}

Finally, owing to the relevance of the $\Omega=1$
degeneracy, we consider the model of $L$, for simplicity equally spaced, 
s.p.l. all having 
$\Omega=1$,  a situation occurring in metals and in deformed nuclei.
In this instance two positive $G_{\rm cr}$ always exist in the lowest s.p.l.
when $L\geq 3$ (in fact $G_{\rm cr}^{(1)-}\to\infty$ for $L=2$).
Moreover a positive $G_{\rm cr}^{(1)-}$ implies complex $E_1$ and $E_2$ for
$G<G_{\rm cr}^{(1)-}$ and since for small $G$ both the pair energies are real,
they should evolve from an unperturbed configuration connected with
the next higher lying s.p.l., as illustrated in fig.\ref{fig:4}.
Numerically we have found, for this model, 
that two $G_{\rm cr}$ appear on the second level when 
$L\ge 9$ and on the third level when $L\ge 16$.
Thus in the $\Omega=1$ model for two pairs to form, so to speak, a 
{\em quartet} it is necessary that they live on adjacent s.p.l. in the
unperturbed configuration. Furthermore the more excited the configuration is
the more not only $G$, but $L$ as well, should be larger for the merging to 
occur, a fact clearly reflecting the competition between the mean field
and the pairing force.
Finally observe that here, at variance
with the finding of ref.~\cite{Roman,Hase,Roman:2002dh},
$G_{\rm cr}^{(\nu +1)+}> G_{\rm cr}^{(\nu)+}$:
this is simply because we measure the s.p.e. from the bottom of the well
rather than from the Fermi surface.

\subsection{The significance of $G_{\rm cr}$}
\label{sec:Gcr}

In \ref{sec:angcr}
we have found that, when a s.p.l. with a degeneracy 
$\Omega=1$ exists, at least two states of the $n=2$ system carry a critical 
value $G_{\rm cr}$. It is then 
natural to ask what is the relationship of the latter 
with the $G_{\rm cr}$ found in \cite{BaCeMoQu-02}
where a system with $n=1$ was addressed. 
Actually in \cite{BaCeMoQu-02} 
$G_{\rm cr}$ marked the transition from a mean field dominated regime
to the one ruled by the pairing interaction (as far as the trapped 
states are concerned). Moreover, in \cite{BaCeMoQu-02} we have proved the
existence of $G_{\rm cr}$ 
for any potential well with s.p.l. of 
increasing degeneracy.

To connect the values of $G_{\rm cr}$ for the systems with 
$n=1$ and $n=2$, respectively, we consider again the harmonic oscillator 
potential. For this well, in the $n=1$ case, by explicitly computing the 
variance and the total degeneracy of the levels, the
critical value of $G$, in units of $\hbar\omega_0$, is
\begin{equation}
\label{gcrho}
G_{\rm cr}\simeq \frac{8}{L^2}
\end{equation} 
when $L$ is large. 
For the $n=2$ system first notice that, at large $L$, in 
$G_{\rm cr}^{(1)\pm}$
the moment ${\cal P}_1$ dominates over ${\cal P}_2$: 
hence, in this condition, 
\begin{equation}
G_{\rm cr}^{(1)+}\simeq G_{\rm cr}^{(1)-}~.
\end{equation}
Moreover, again for $L$ large enough, 
\begin{equation}
{\cal P}_1\simeq \frac{L^2}{8}
\end{equation}
from where the equality
\begin{equation}
G_{\rm cr}(n=1)=G_{\rm cr}(n=2)
\end{equation}
follows in the asymptotic $L$ limit. 
From this outcome it follows that also when $n=2$ the relevant dynamical 
element for $G<G_{\rm cr}^{(1)+}$ is the mean field, whereas for 
$G>G_{\rm cr}^{(1)-}$ is the pairing force, as far as the system's ground 
state is concerned. 

One might ask how this interpretation can be reconciled with the findings 
related to the potential well of equally spaced s.p.l. 
with $\Omega=1$, for which no $G_{\rm cr}$  occurs in the $n=1$ 
system, whereas a $G_{\rm cr}^{(+)}$ does exist in the $n=2$
system. 

To understand this occurrence consider the structure of the trace of
$\hat H_{pair}$.
This, in fact, for the $n=1$ system and {\em as far as the trapped modes
are concerned} is found, in the strong coupling limit, to read
\begin{equation}
\label{eq:Sigg}
\Sigma^{(\gamma)}=\frac{1}{L}\sum_{k=2}^L z_k
=\frac{1}{\gamma+2}~,
\end{equation}
where $z_k=(E_k-2e_{k-1})/(2e_k-2e_{k-1})$ actually
measures the shift of each trapped pair energy $E_k$ $(k=2,\cdots L)$ 
from the corresponding unperturbed energy $2e_k$.
The above holds for any well with $L$ (large) equally spaced s.p.l. with 
degeneracy
\begin{equation}
\Omega_k^{(\gamma)} \propto k^\gamma\,.
\end{equation}
We thus see that (\ref{eq:Sigg}) yields
\begin{equation}
\Sigma^{(h.o.)}=\frac{1}{4}
\end{equation}
for the harmonic oscillator ($\gamma=2$)
and
\begin{equation}
\Sigma^{(\Omega_k=1)}=\frac{1}{2}
\end{equation}
for the $\Omega=1$ model ($\gamma=0$).

Hence for the latter the pairing force is so strongly active among
the trapped pair energies to prevent the occurrence of
a transition from the pairing to the mean field dominated physics, 
whereas the opposite happens for the harmonic oscillator well.

In addition the trace of $\hat H_{pair}$ entails that in the $\Omega=1$ model
the collective mode is weaker than in the harmonic oscillator case: indeed to 
set it up not only a strong $G$, but a large degeneracy as well, is needed.

Concerning the $n=2$ case in the $\Omega=1$ model, the pairing 
dominance regime, defined by $G>G_{\rm cr}^{-}$, is indeed postponed at
values of $G$ larger than in the harmonic oscillator case.
In fact one deduces from
\eqref{eq:29n}, in the large $L$ limit, the expression 
\begin{equation}
G_{\rm cr}\simeq\frac{2}{\log L}~,
\label{gcrito1}
\end{equation}
which, {\em en passant}, fixes the domain of validity 
of the perturbative expansion in $G$~\cite{Sch01} for the $\Omega=1$ model.

One might accordingly conjecture that no transition between different
regimes will occur in this case among the trapped energies: we are
currently performing the analysis of the behavior of the $z_k$ versus
$G$ for the $v_l=2$ states to ascertain whether this statement holds true.

\subsection{Comparing the Bogoliubov quasi-particle and the 
Richardson's exact solution}
\label{sec:BCS}

In this Section we compare the exact solutions of the Richardson equations
with the BCS solution in terms of Bogoliubov quasi-particles.
To this scope we self-consistently solve the well-known BCS equations for $N$
fermions living in $L$ levels, namely
\begin{eqnarray}
&&v_\nu^2 = \frac{1}{2} \left[1-
\frac{\epsilon_\nu-\lambda}{\sqrt{(\epsilon_\nu-\lambda)^2+\Delta^2}}\right]
= 1- u_\nu^2\,,
\\
&&\sum_{\nu=1}^L 
\frac{\Omega_\nu}{\sqrt{(\epsilon_\nu-\lambda)^2+\Delta^2}}=\frac{2}{G}\,,
\label{lam}
\\
&&\sum_{\nu=1}^L 
\Omega_\nu \left[1-
\frac{\epsilon_\nu-\lambda}{\sqrt{(\epsilon_\nu-\lambda)^2+\Delta^2}}\right]=N
\end{eqnarray}
being $\Delta= G\sum_\nu\Omega_\nu u_\nu v_\nu$ the gap, 
$\lambda$ the chemical potential and 
$\epsilon_\nu= e_\nu-G v_\nu^2$.

In the BCS framework 
the excitation energy of a system with seniority $v$ is given 
by the energy of $v$ quasi-particles, each 
carrying an energy
\begin{equation}
E_\nu^{QP}=\sqrt{(\epsilon_\nu-\lambda)^2+\Delta^2}\,.
\label{eq:eQP}
\end{equation}

\begin{figure}[ht]
  \begin{center}
\def\epsfsize#1#2{0.9#1}
\epsfbox[170 520 450 820]{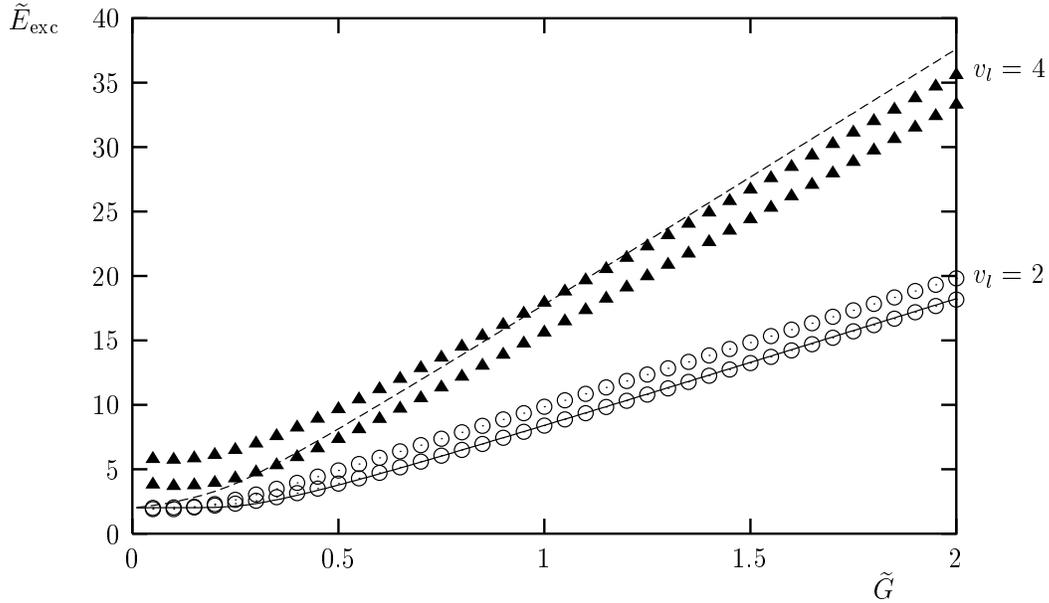}
    \caption{Behaviour with $\tilde G$ of the excitation energy
of a two-pairs system in a harmonic oscillator well with $L=3$.
The exact solutions with $v_l=2$ (circles) and $v_l=4$ (triangles) are compared
with the BCS results for $v=2$ (solid line) and $v=4$ (dashed line).}
    \label{fig:8}
  \end{center}
\end{figure}

In fig.~\ref{fig:8} we display and compare 
the exact excitation energies $E_{\rm exc}=E(v_l)-E(g.s.)$ for a $L=3$ 
harmonic oscillator well and 
the corresponding Bogoliubov's quasi-particles predictions for a
$v=2$ and a $v=4$ state, whose excitation energies are
$2 E_1^{QP}$  and $2 (E_1^{QP}+ E_2^{QP})$, respectively.
It appears that 
for $\tilde G$ larger than the highest critical value ($\tilde G_{\rm cr}^-
\simeq 0.6$, see fig.~1) both the BCS and the exact excitation energies become
linear functions of $\tilde G$ and, remarkably, 
are very close to each other, in particular for 
$v_l=2$. This result on the one hand 
shows that the BCS theory is a valid (at least for the lowest excitations)
approximation of the exact physics 
even when the number of pairs and levels is indeed low, 
thus strengthening the correspondence between seniority
and ``like-seniority''.
On the other hand it confirms that the Richardson exact 
solutions behave like
(\ref{eq:eQP}) 
in the strong $G$ limit, as found by Gaudin~\cite{Gau95},
who proved that for large $L$ the excitation energies are given indeed by
sums of terms like (\ref{eq:eQP}).
It is remarkable that this appears to be approximately true already for
$L=3$.

Finally we like to remark that the exact energies displayed in fig.~\ref{fig:8}
correspond, microscopically, to two particles-two holes (the $v_l=2$) and to
four particles-four holes (the $v_l=4$) excitations without the breaking
of any pair.
In other words they are associated with the promotion of one or two pairs
to higher lying s.p.l.

It is also worth mentioning that while the s.p.e.
$\epsilon_\nu$ are almost constant in $G$, 
the quasi-particle energies $E_\nu^{QP}$ grow linearly with the latter
(for $G>G_{\rm cr}$) and their spreading in energy is lower 
than the one among  the $\epsilon_\nu$. 

We finally note that it would be interesting to compare the
Richardson exact result with the energy of the excited $0_2^+$ state 
known as ``pairing vibrational'' state calculated in the quasi-particle 
representation. This we leave for future work.

\section{The wave functions}
\label{sec:wf}

In this Section we examine the wave functions of the states so far discussed.

As well-known, 
the $v=0$ eigenfunctions of $\hat H_{pair}$ for a $n$ pairs system 
are expressed in 
terms of the collective Grassmann variables $\Phi_{\mu}$ (referred to as 
$s$-quasibosons), 
according to~\cite{Barbaro:1999fh}
\begin{equation}
\label{eq:cx1}
\psi_n[\Phi^*](m) = \prod_{k=1}^n {\cal B}^*_k(m)
\end{equation}
where the set of indices $(m_1,\dots,m_n)$ labeling the 
unperturbed configuration from where the state develops are collectively 
denoted by $m$.
Since
\begin{equation}
{\cal B}^*_k(m) = \sum_{\mu=1}^L \beta_\mu^{(k)}(m)
 \Phi^*_\mu 
\label{psin}
\end{equation}
can be viewed as the wave function of a pair, \eqref{eq:cx1} actually 
corresponds to the wavefunction of $n$ independent pairs.
The $\beta$ coefficients are related to the eigenvalues $E_k$ according to:
\begin{equation}
  \label{eq:az12}
  \beta_\mu^{(k)}(m)=\frac{C_k(m)}{2 e_\mu-E_k(m)}
\end{equation}
the $C_k(m)$ being normalisation factors.
When no confusion arises the index $(m)$ will be dropped.

Since $(\Phi_\mu,\Phi_\nu^*)=\Omega_\mu\delta_{\mu\nu}$, 
it is convenient to replace the $\Phi$'s and the $\beta$'s with
\begin{equation}
  \label{eq:ac4}
  \tilde\Phi_\mu^*=\frac{\Phi_\mu^*}{\sqrt{\Omega_\mu}}\qquad\qquad
\mbox{and}\qquad\qquad  
\tilde\beta_\mu^{(k)}=\sqrt{\Omega_\mu}\beta_\mu^{(k)}~,
\end{equation}
respectively.
The normalisation of the state $|(m)>$ then requires
\begin{equation}
  \label{eq:ax58}
   C_k(m)=\frac{1}{\sqrt{\sum_\mu\dfrac{\Omega_\mu}{|2e_\mu-E_k(m)|^2}}}\,.
\end{equation}
In the one pair case the above with $k=1$ indeed entails $<(m)|(m)>=1$, whereas
for a two pairs state 
(\ref{eq:ax58}) can still be used (of course with $k=1$ for the first
and $k=2$ for the second pair energy),
but the associated state is no longer normalised to 1.

In \cite{BaCeMoQu-02} we have investigated the wave functions of a 
single pair living in many levels. We now study the behaviour with $G$ of the 
wave functions of two pairs, our interest being focused on the states 
undergoing a transition, in the sense discussed in Sec.~\ref{sec:crit}.

\begin{figure}[htp]
  \begin{center}
    \mbox{
      \begin{tabular}{cc}
        \epsfig{file=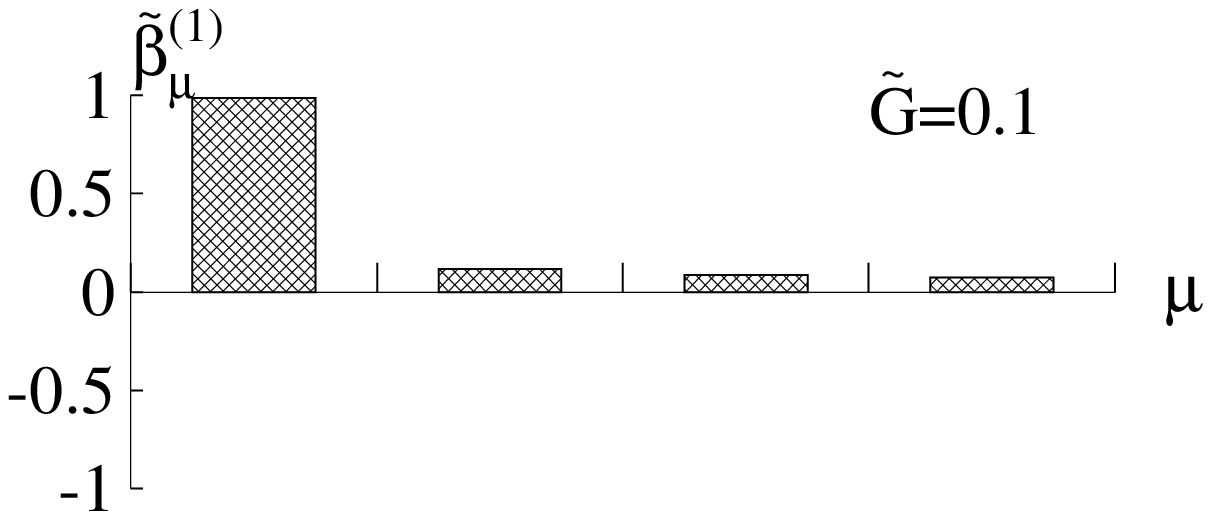,width=6cm,height=2cm}&
        \epsfig{file=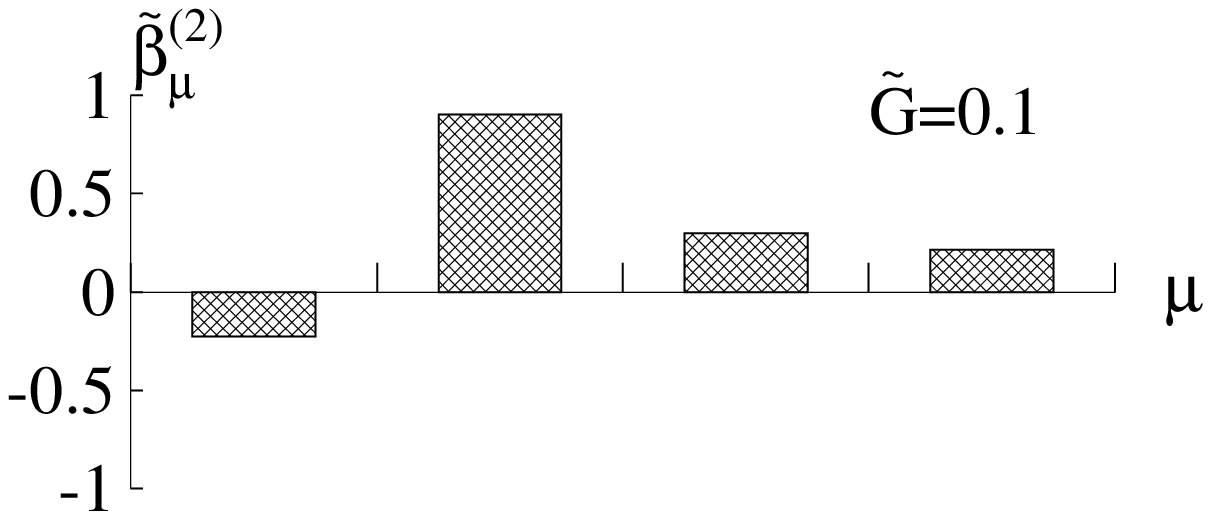,width=6cm,height=2cm}
        \\
        \epsfig{file=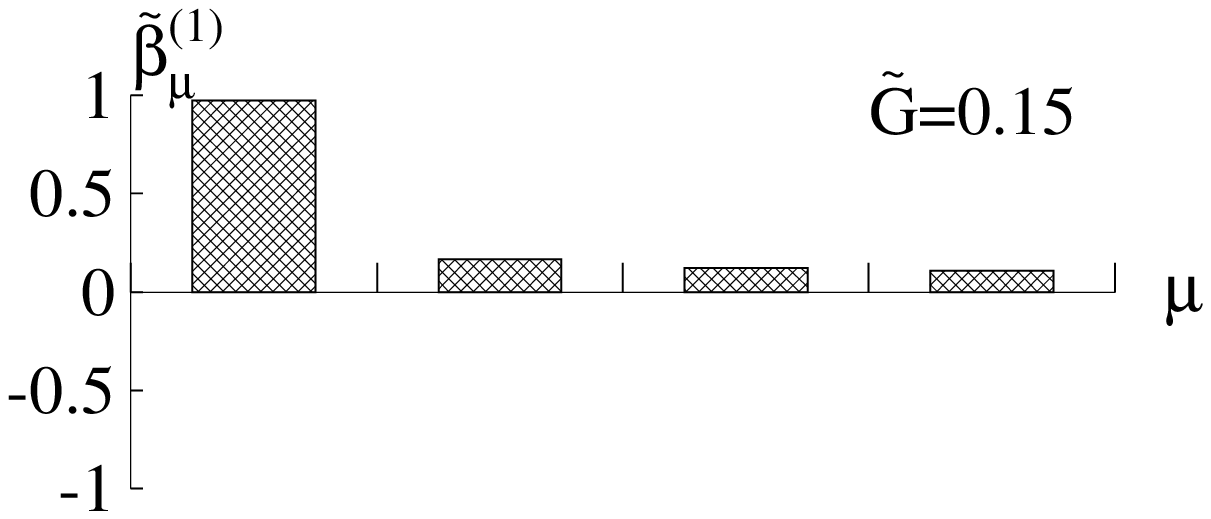,width=6cm,height=2cm}&
        \epsfig{file=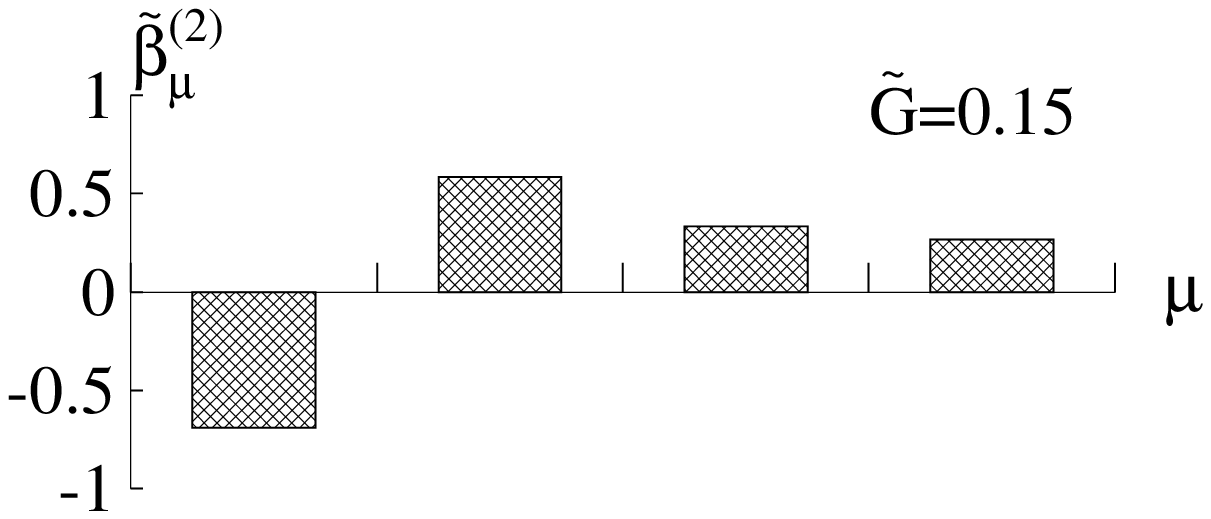,width=6cm,height=2cm}
        \\
        \epsfig{file=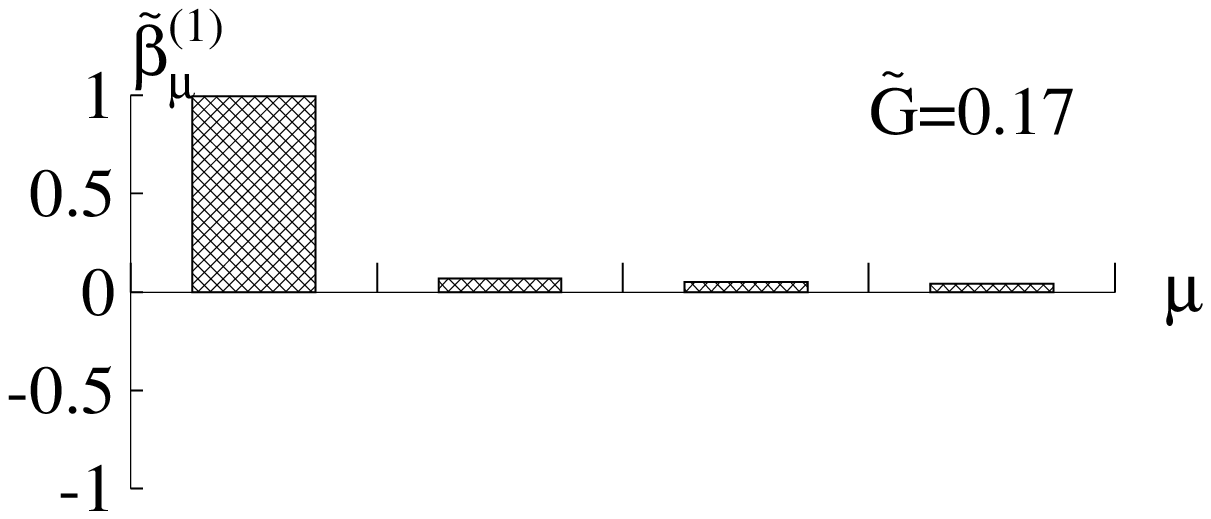,width=6cm,height=2cm}&
        \epsfig{file=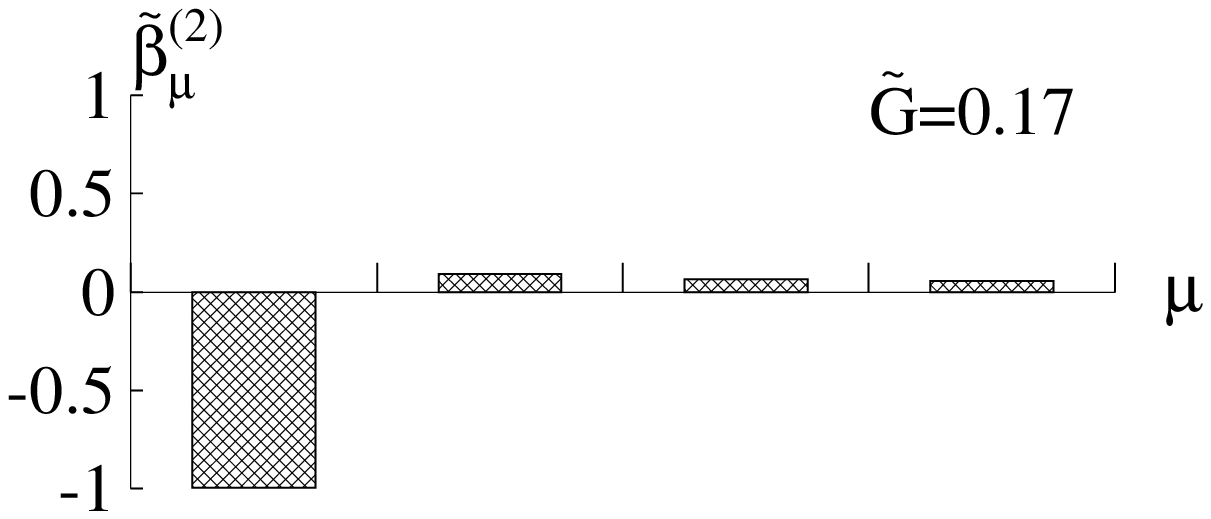,width=6cm,height=2cm}
        \\
        \epsfig{file=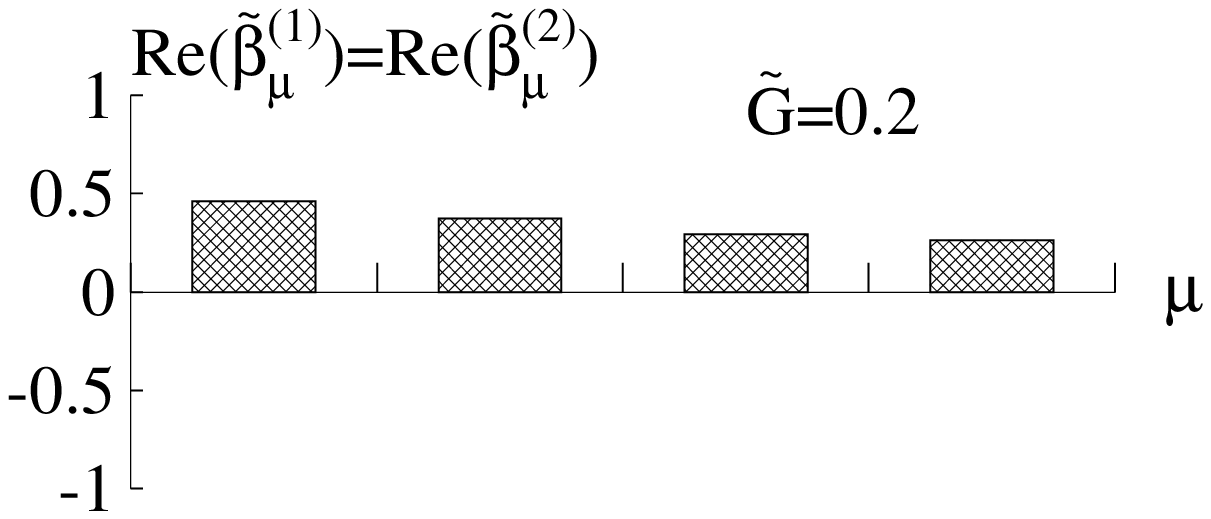,width=6cm,height=2cm}&
        \epsfig{file=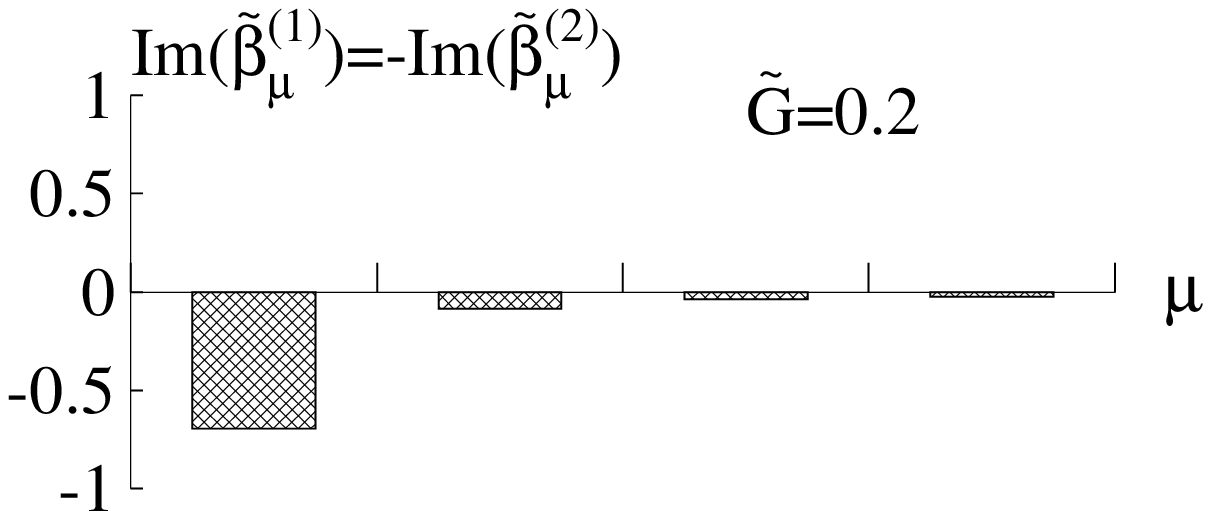,width=6cm,height=2cm}
        \\
        \epsfig{file=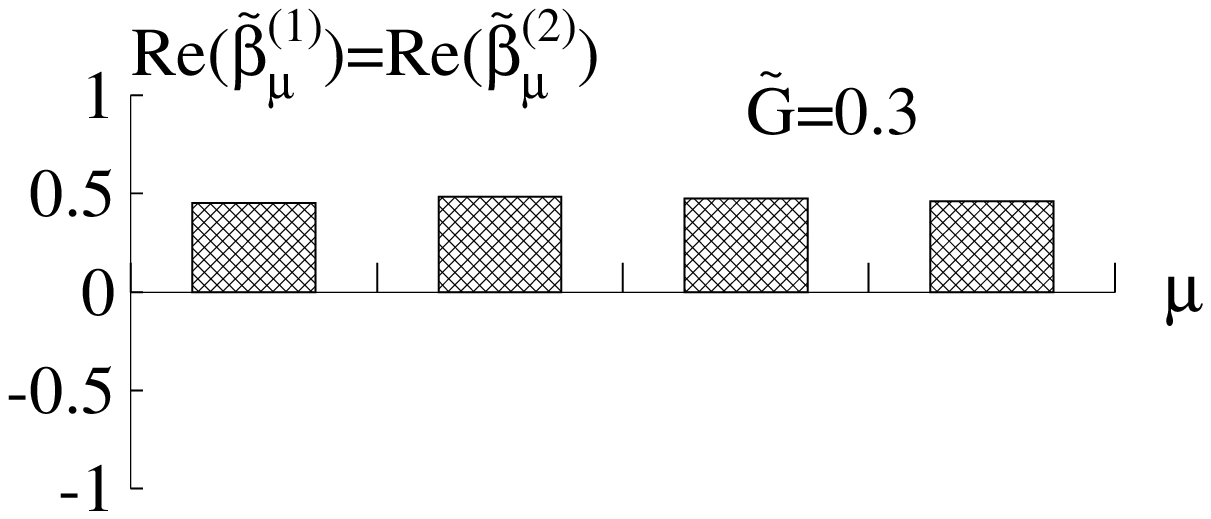,width=6cm,height=2cm}&
        \epsfig{file=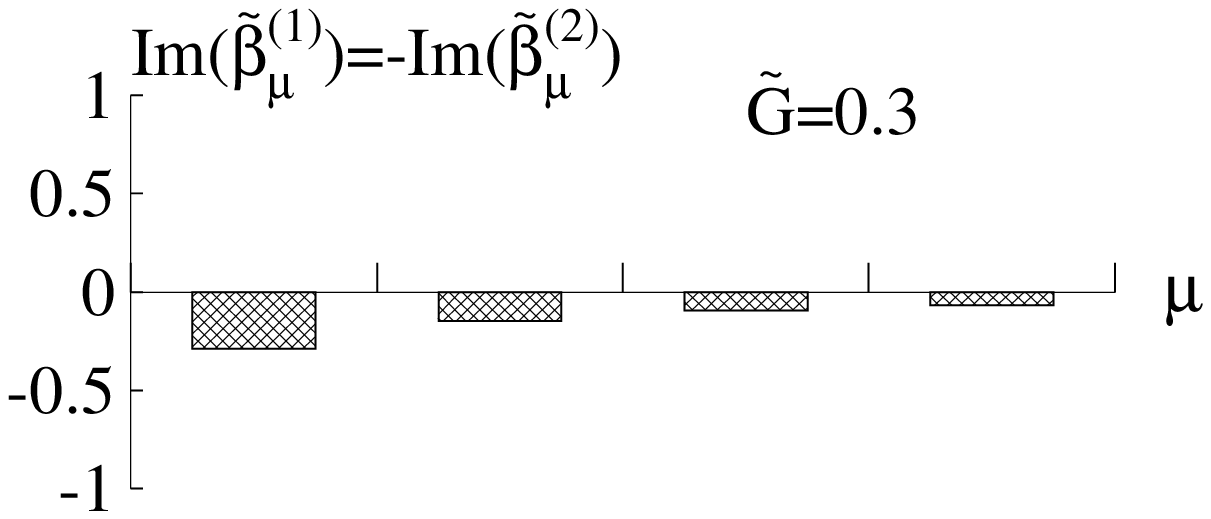,width=6cm,height=2cm}
        \\
        \epsfig{file=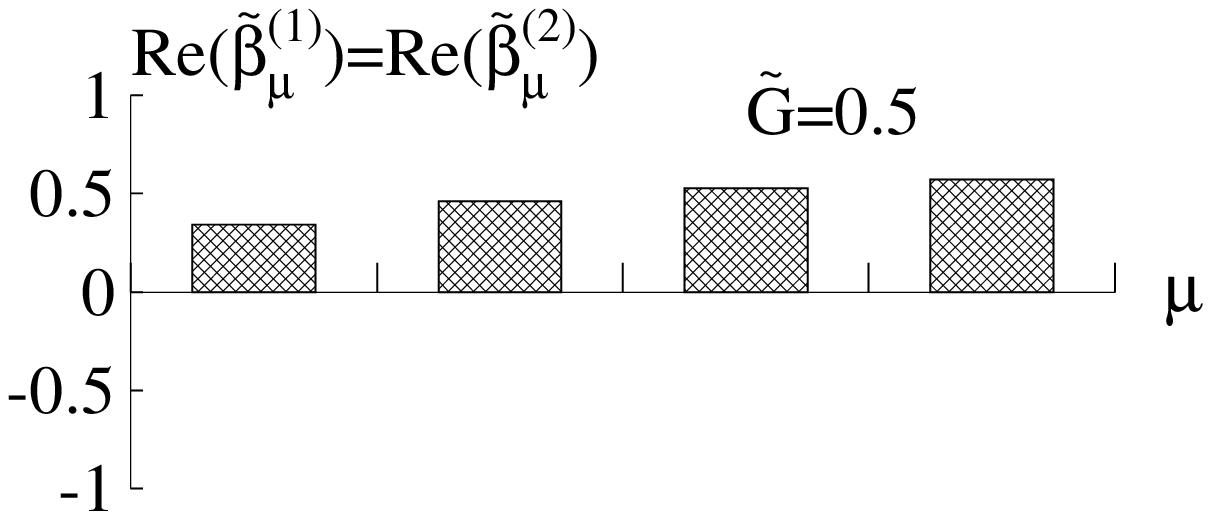,width=6cm,height=2cm}&
        \epsfig{file=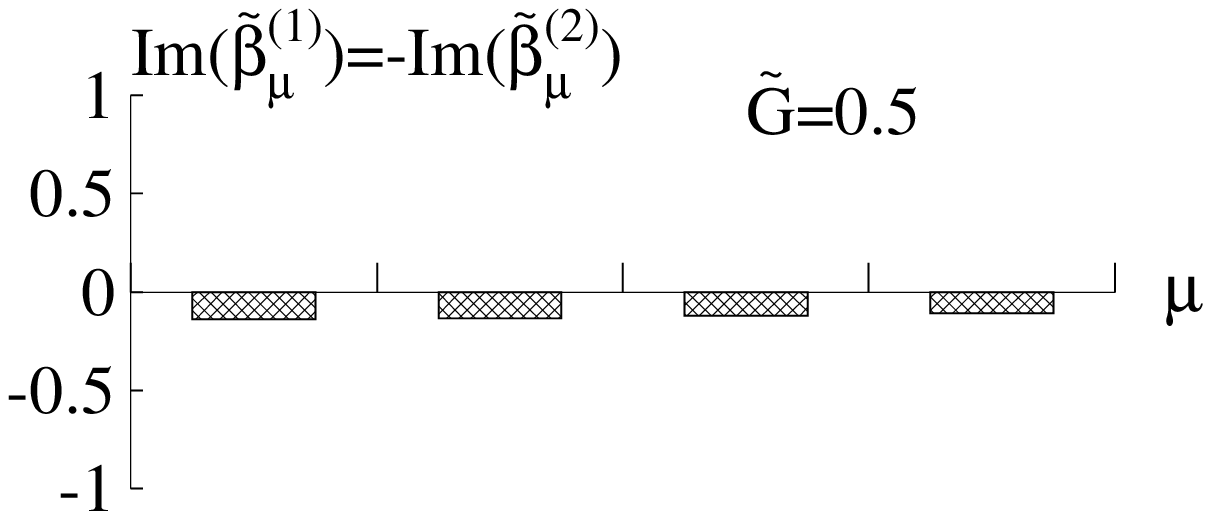,width=6cm,height=2cm}
        \\
        \end{tabular}
        }
    \caption{In the $L=4$ levels harmonic oscillator well, we display for 
      different values of $\tilde G$ the coefficients 
    $\widetilde\beta_\mu^{(k)} $ (or their real and imaginary part) as 
     functions of the index $\mu $ labeling the s.p.l. for the $v_l=0 $ 
     state. The index $k$ labels the pairs.}
    \label{fig:5}
  \end{center}
\end{figure}

\begin{figure}[htp]
  \begin{center}
    \mbox{
      \begin{tabular}{cc}
        \epsfig{file=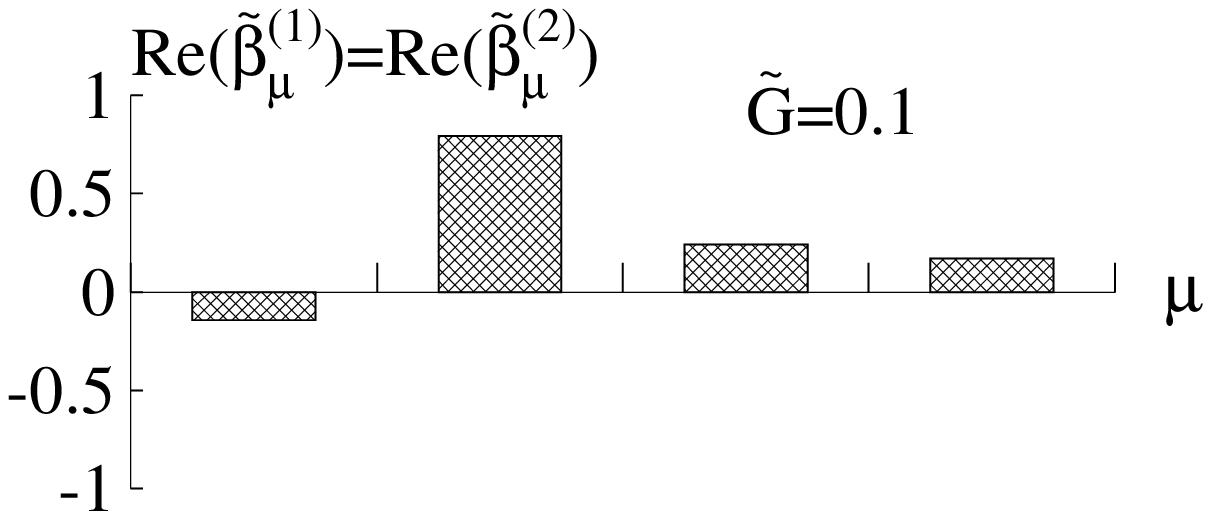,width=6cm,height=2cm}&
        \epsfig{file=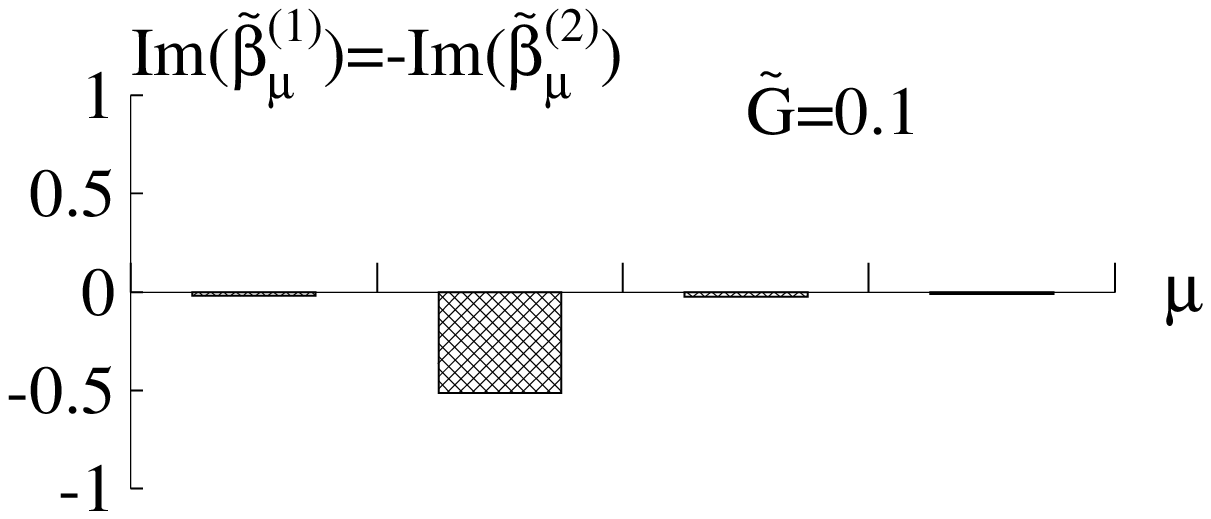,width=6cm,height=2cm}
        \\
        \epsfig{file=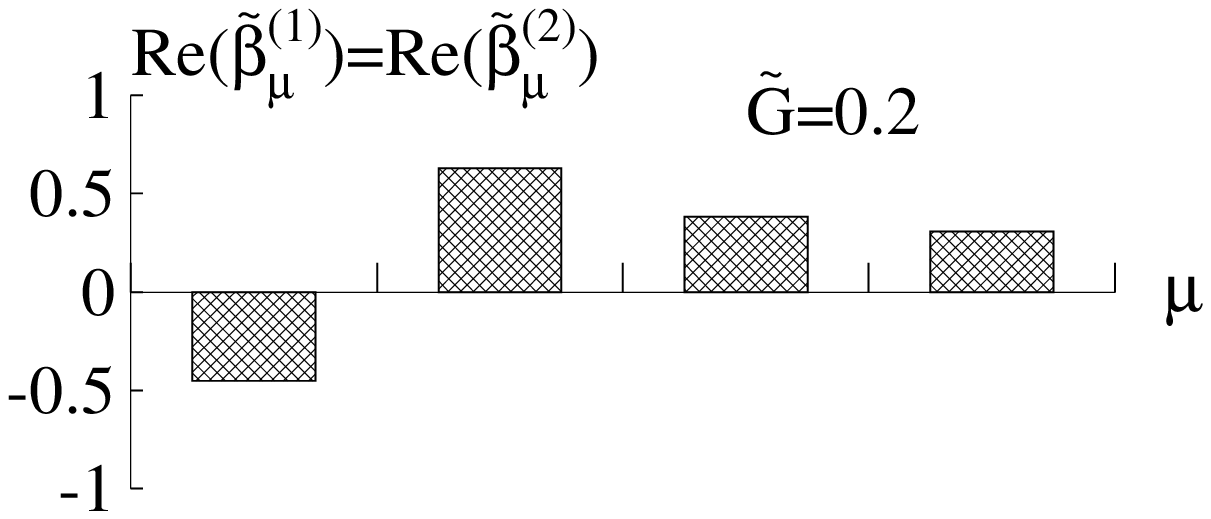,width=6cm,height=2cm}&
        \epsfig{file=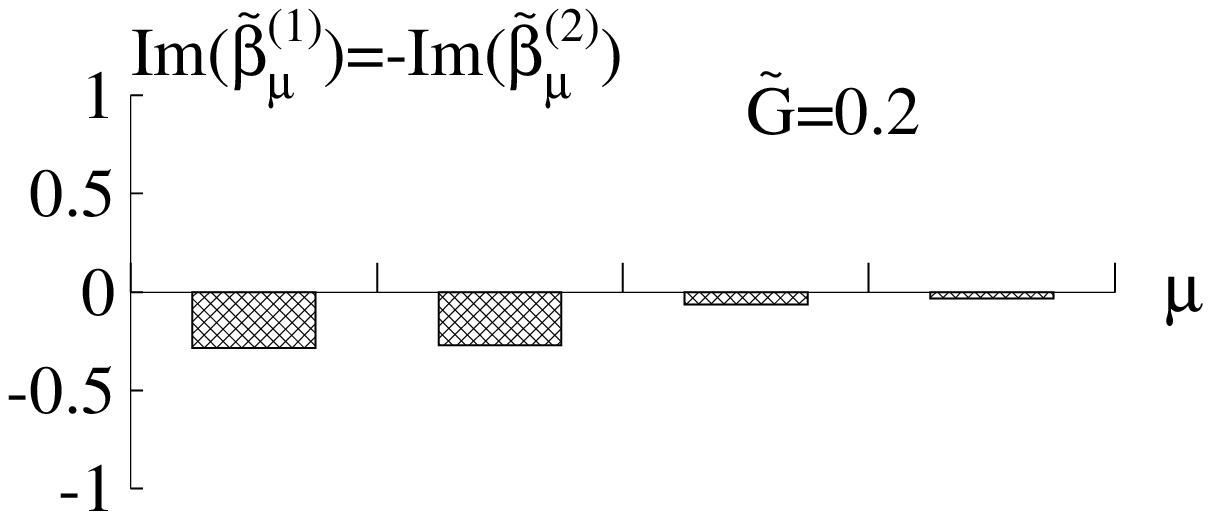,width=6cm,height=2cm}
        \\
        \epsfig{file=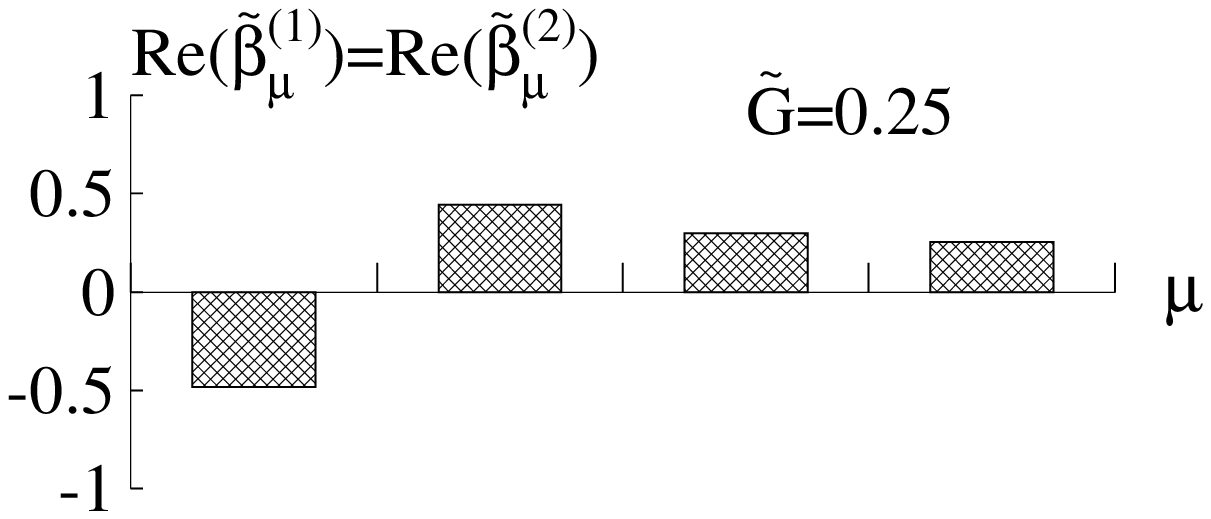,width=6cm,height=2cm}&
        \epsfig{file=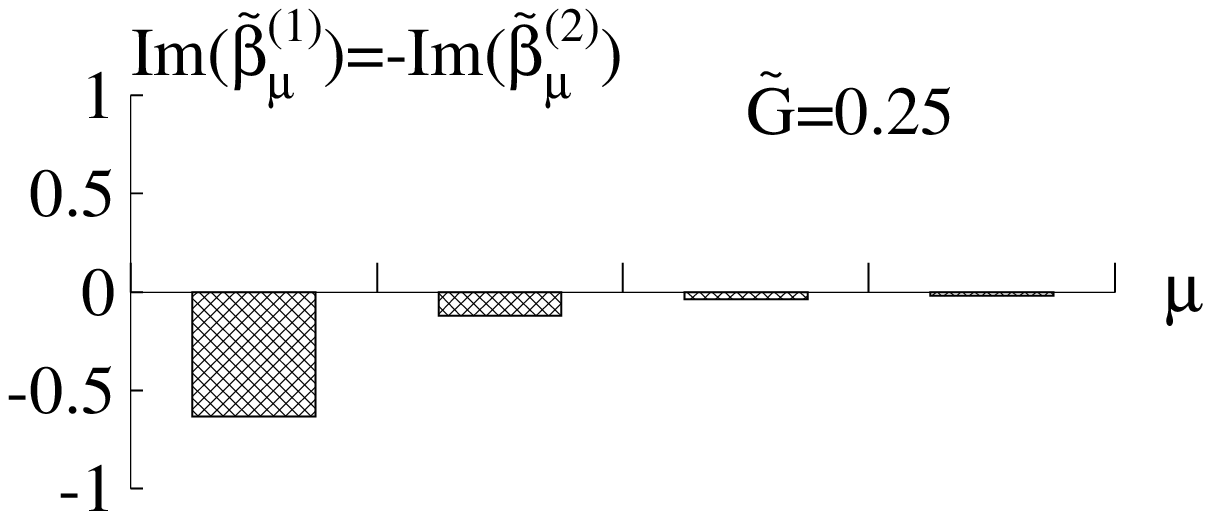,width=6cm,height=2cm}
        \\
        \epsfig{file=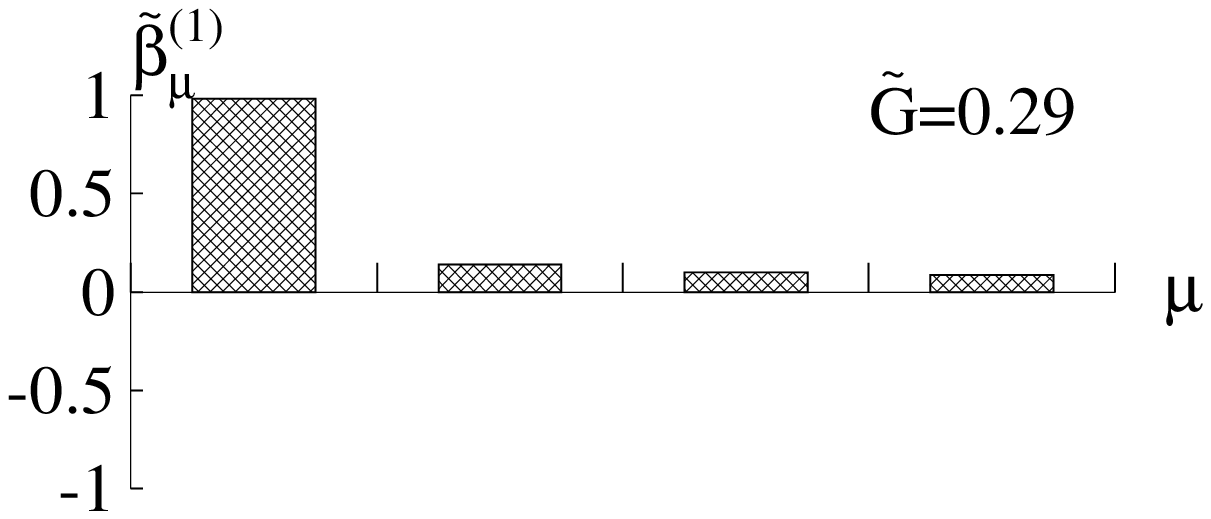,width=6cm,height=2cm}&
        \epsfig{file=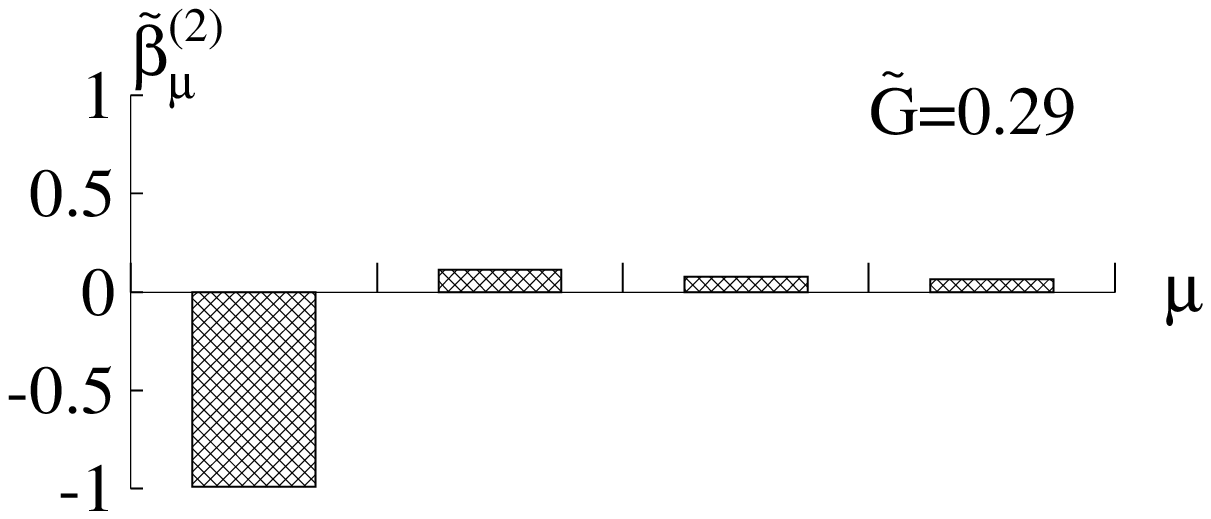,width=6cm,height=2cm}
        \\
        \epsfig{file=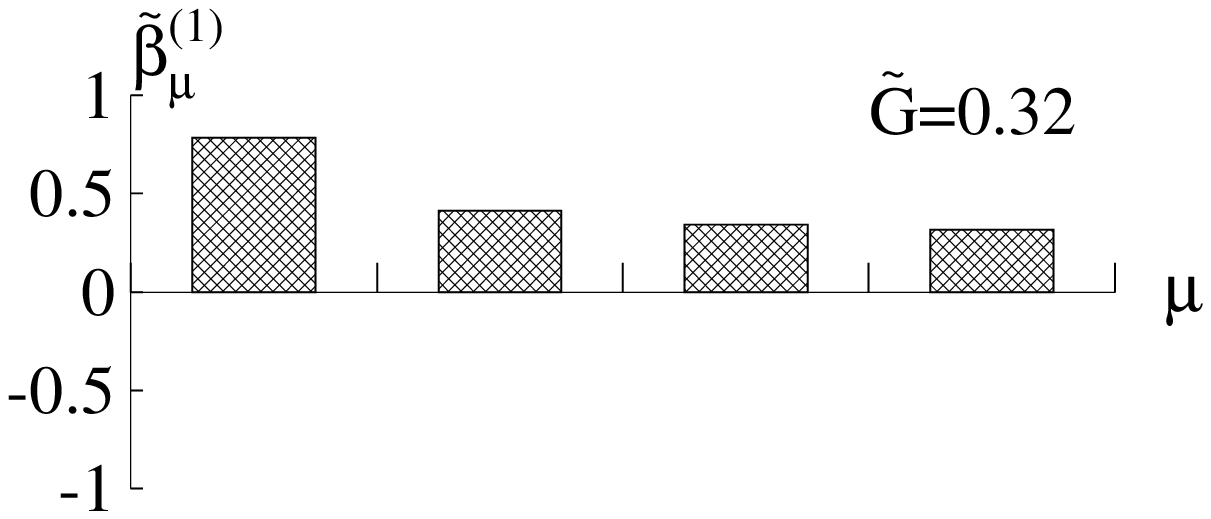,width=6cm,height=2cm}&
        \epsfig{file=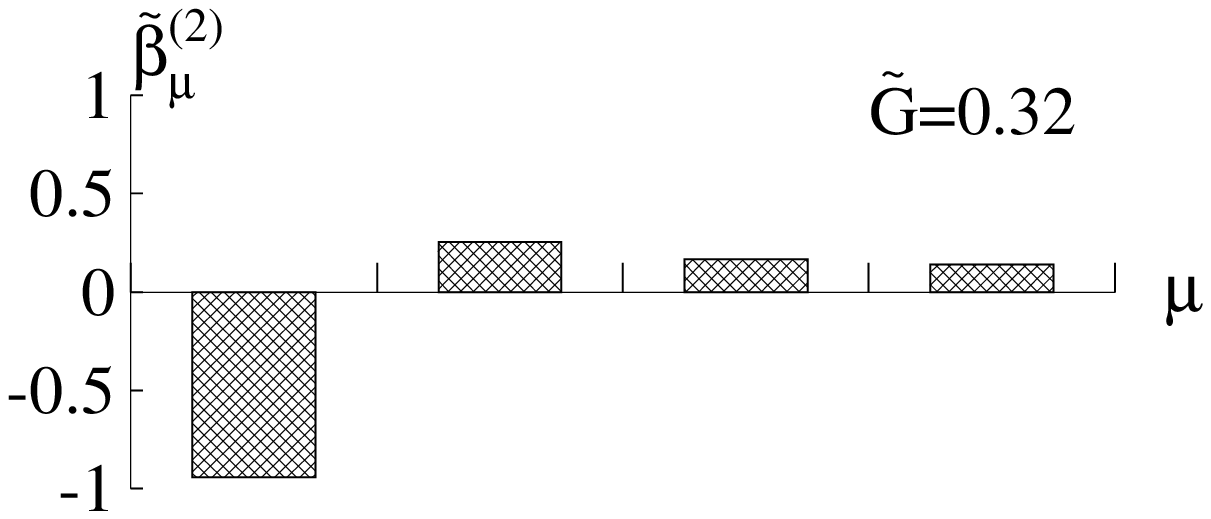,width=6cm,height=2cm}
        \\
        \epsfig{file=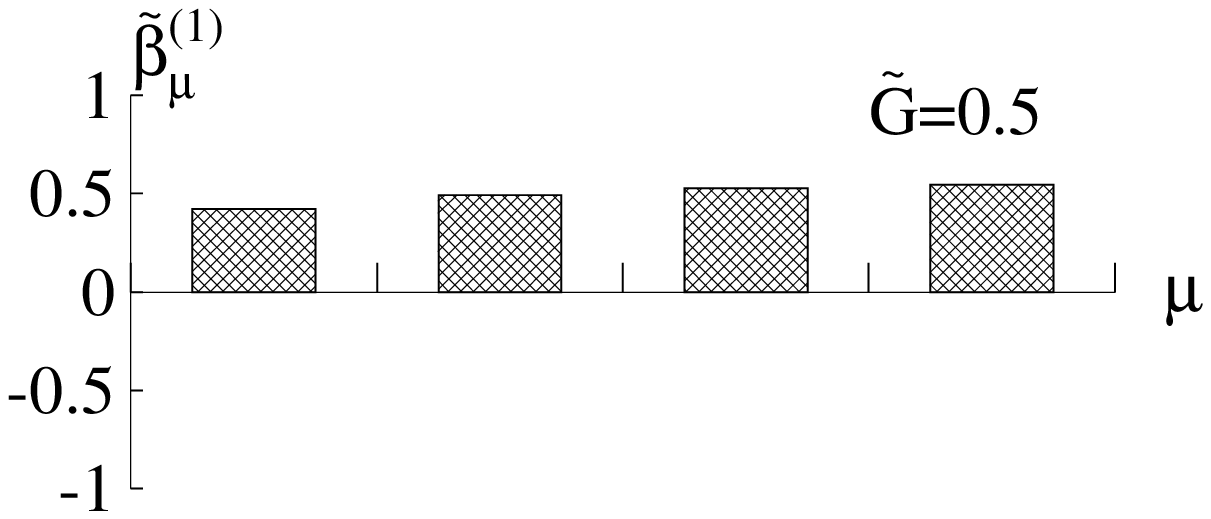,width=6cm,height=2cm}&
        \epsfig{file=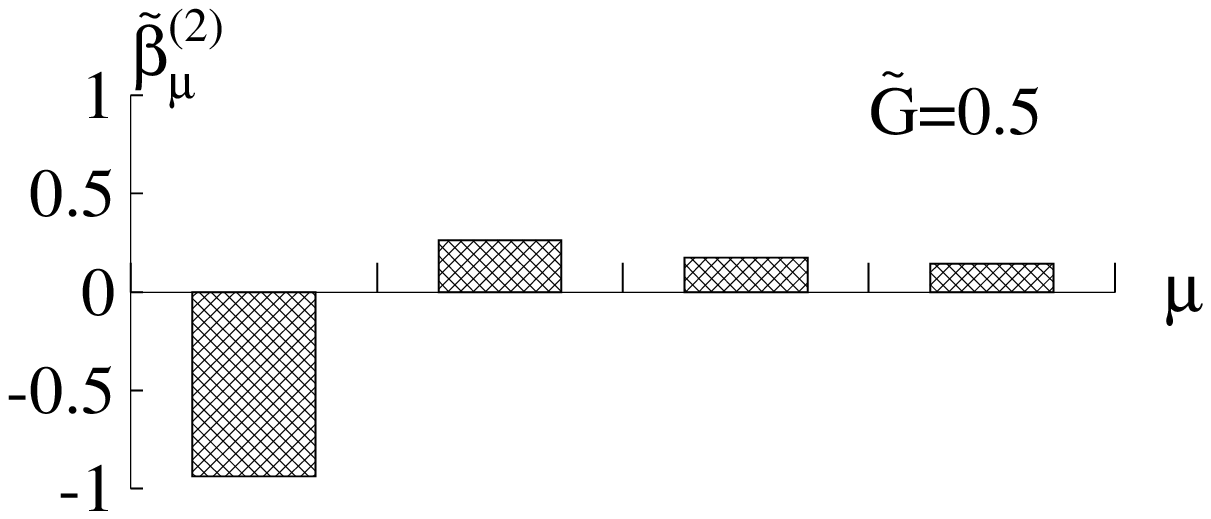,width=6cm,height=2cm}
        \end{tabular}
      }
    \caption{The same as in fig.~\protect\ref{fig:5}, but for the $v_l=2$ 
    state, made up of two pairs living on the second s.p.e. level. }
    \label{fig:6}
  \end{center}
\end{figure}

For sake of illustration we take four s.p.l. of a harmonic oscillator well 
and consider the states $v_l=0$ (namely the configuration $(m_1,m_2)=(1,2)$)
and one of the $v_l=2$ (associated with the configuration $(m_1,m_2)=(2,2)$).
We display in fig.~\ref{fig:5} and fig.~\ref{fig:6} respectively, for different
values of $\tilde G$, their coefficients $\widetilde\beta_\mu^{(k)}$ 
as functions of the index $\mu$. 

In the $v_l=0$ case, for $\tilde G< \tilde G^{(1)+}_{\rm cr}$, the coefficients
$\widetilde\beta_\mu^{(k)}(1,2) $ are real, while for 
$\tilde G> \tilde G^{(1)+}_{\rm cr}$, 
they become complex conjugate (hence we display 
both their real and the imaginary part); the opposite occurs in the 
$v_l=2$ case.      

As it appears in 
fig.~\ref{fig:5}, the $v_l=0$ state initially has essentially
one pair in the first and one in the second s.p.l. As $\tilde G$ approaches 
$\tilde G^{(1)+}_{\rm cr}=0.171$, both pairs almost
sit on the first level (thus displaying an apparent
violation of the Pauli principle, which actually becomes a true violation
at the critical point), where however the system cannot live.
Finally for $\tilde G\gg \tilde G^{(1)+}_{\rm cr}$, 
the weight of all the components of $\widetilde\beta^{(1)}(1,2)$ and  
$\widetilde\beta^{(2)}(1,2)$ are almost the same, 
thus displaying a collective behaviour. 

The (2,2) $v_l=2$ state (see fig.~\ref{fig:6}) instead starts
up with two pairs on the second s.p.l. (see the case $\tilde G=0.1$
in fig.~\ref{fig:6}).
As $\tilde G$ approaches $\tilde G^{(1)-}_{\rm cr}=0.287$, 
the two pairs seem to live on the lowest single particle level
(again forbidden by the Pauli principle).
Finally for $\tilde G\gg \tilde G^{(1)-}_{\rm cr}$, all the components of 
$\widetilde\beta^{(1)}(2,2)$ have almost the same weight (like the components 
of the collective state of one pair) while only the first component of 
$\widetilde\beta^{(2)}(2,2)$ is significant (like for the trapped state of one 
pair). 

In conclusion we provide analytic expressions for the 
$\tilde \beta^{(k)}_\mu$ coefficients for
$G\to0$, $G\to\infty$ and $G=G_{\rm cr}$.

The weak coupling limit is immediate and
one obtains $\tilde \beta^{(1)}_\mu(m_1,m_2)=\delta_{m_1\mu}$ and 
$\tilde \beta^{(2)}_\mu(m_1,m_2)=\delta_{m_2\mu}$.

The strong coupling limit is easily handled only when
$v_l=0$ (independently of $\Omega_1$), when the s.p.e. are small
with respect to $E_1$, $E_2$. In fact one then gets
\begin{equation}
  \label{eq:xa1}
  \tilde\beta^{(1,2)}_\mu=\frac{\sqrt{\Omega_\mu}}{\Omega}
  \left(\sqrt{\Omega-1}\pm i\right)~.
\end{equation}

The structure of the wave function at the critical points (assuming 
the transition to occur on the $\mu=1$ level) is more delicate.
To the order $x^2$ one finds 
\begin{eqnarray}
  \label{eq:ab1}
  \tilde\beta^{(1)}_1&=&-\left\{1-\frac{x^2}{2}
    {\cal P}_{(2)1}^2\right\}~,\\
  \tilde\beta^{(2)}_1&=&+\left\{1-\frac{x^2}{2}
    {\cal P}_{(2)1}^2\right\}~,\\
  \tilde\beta^{(1)}_{\mu\not=1}&=&\sqrt{\Omega_\mu}\left\{
    \frac{x}{2\epsilon_\mu
      -2\epsilon_1}+\left(\frac{x}{2\epsilon_\mu
        -2\epsilon_1}\right)^2\right\}\\
  \tilde\beta^{(2)}_{\mu\not=1}&=&\sqrt{\Omega_\mu}\left\{
    \frac{x}{2\epsilon_\mu
      -2\epsilon_1}-\left(\frac{x}{2\epsilon_\mu
        -2\epsilon_1}\right)^2\left[1\pm 2{\cal P}_{(2)1}
    (2\epsilon_\mu      
         -2\epsilon_1)\right]\right\}~.\nonumber
\label{55}
\\~
\end{eqnarray}
In the above $x$ is connected to $G$ by the relation 
\begin{equation}
  \label{eq:31n}
  x^2=\mp 2
\left(G-G_{\rm cr}\right)
{\cal P}_{(2)1}\frac{\left({\cal P}_{(1)1}
      \pm{\cal P}_{(2)1}\right)^2}
  {3{\cal P}_{(2)1}^4+4{\cal P}_{(2)1}{\cal P}_{(3)1}^3
    +{\cal P}_{(4)1}^4
    }+ {\cal O}\left(\left(G-G_{\rm cr}\right)^2\right)~,
\end{equation}
and in \eqref{55} the double sign refers to the two critical points 
$G_{\rm cr}^{(1)\pm}$. 
Notice that the 
$\tilde\beta^{(1,2)}_{\mu\not=1}$ start linearly in $x$, i.e., in
$\sqrt{G-G_{\rm cr}}$, thus displaying a branch point in the control parameter
$G$ at its critical value.

\section{The transition amplitudes}
\label{sec:tran}

In this Section we study the pair transfer
amplitudes from the vacuum to a one pair state and from a one pair to a two 
pairs state as a function of  $G$, to explore how
the transition between the two different regimes previously discussed is 
reflected in the matrix element. 

We thus study the transition amplitude induced by the operator
\begin{equation}
  \label{eq:ay1}
  \hat A^\dagger=\sum_\mu \hat A^\dagger_\mu\equiv\sum_{j_\mu}
  \sum_{m_\mu=\frac{1}{2}}^{j_\mu}
  (-1)^{j_\mu+m_\mu}\hat a^\dagger_{j_\mu,m_\mu}\hat a^\dagger_{j_\mu,-m_\mu}
~,\end{equation}
which enters into $\hat H_{pair}$, and examine the matrix 
elements
\begin{equation}
<\text{1~pair}|\hat A^\dagger|0>\ \ \ \mbox{and}\ \ \ 
<\text{2~pairs}|\hat A^\dagger|\text{1~pair}>~.
\label{eq:tme}
\end{equation}

In the fully degenerate case the general expression for these transition
amplitudes between states with any number $n$ of pairs and seniority $v$,
can be obtained by inserting into the pairing spectrum a complete 
set of states $|n',v'>$. Indeed one has
\begin{equation}
\begin{split}
\sum_{n',v'} & <n+1,v|\hat A^\dagger|n',v'> 
<n',v'|\hat A|n+1,v>
\\&
=\left|<n+1,v|\hat A^\dagger|n,v>\right|^2
=\left(n+1-\frac{v}{2}\right)
    \left(\Omega-n-1-\frac{v}{2}\right)\,,
  \label{eq:degsp1}
\end{split}
\end{equation}
thus getting for the transition amplitude
\begin{equation}
  \label{eq:ax3}
  <n+1,v|\hat A^\dagger|n,v>=\sqrt{\left(n+1-\frac{v}{2}\right)
    \left(\Omega-n-\frac{v}{2}\right)}\,,
\end{equation}
which, while more transparent, 
coincides with the one obtaiend in ref.~\cite{Bes-Broglia}.
Note the $G$-independence of \eqref{eq:ax3}, which, moreover, is diagonal
in the seniority quantum number.

For large $G$ we could expect the matrix elements \eqref{eq:tme} to display
an asymptotic behaviour coinciding with \eqref{eq:ax3}, namely
\begin{eqnarray}
  \label{eq:ax11}
  <1,v_l=0|\hat A^\dagger|0>~&\xrightarrow[G\to\infty]{}&~
  \sqrt{\Omega}\\
  <2,v_l|\hat A^\dagger|1,v_l>~&\xrightarrow[G\to\infty]{}&~
  \begin{cases}
    \sqrt{2(\Omega-1)}&\text{for~}v_l=0\\
    \sqrt{\Omega-2}&\text{for~}v_l=2
  \end{cases}
\end{eqnarray}
if $v_l$ is conserved (or if the above matrix elements are diagonal in $v_l$.) 

However, the number $v_l$ is not
conserved at finite $G$: only in the strong coupling regime the 
vanishing of the matrix element between states of different Gaudin numbers
occurs.

Hence we now compute the transition amplitudes at finite $G$.
For this purpose we first consider the transition from the vacuum to the one 
pair state, namely
\begin{equation}
  \label{eq:az11}
  <(m)|\hat A^\dagger|0>=\sum_\mu\sqrt{\Omega_\mu}\tilde\beta_\mu(m)~,
\end{equation}
which, using \eqref{eq:az12} and the Richardson's equations
for one pair, namely
\begin{equation}
  \label{eq:az20}
  \sum_\mu\frac{\Omega_\mu}{2e_\mu-E}=\frac{1}{G}~,
\end{equation}
can be recast as follows
\begin{equation}
  \label{eq:az13}
  <(m)|\hat A^\dagger|0>=C_1(m)\sum_\mu\frac{\Omega_\mu}{2e_\mu-E(m)}
  =\frac{C_1(m)}{G}~,
\end{equation}
where the label $(m)$ identifies the unperturbed
configuration of the state.

The strong coupling behaviour of \eqref{eq:az13} is different in the
$v_l=0$ and the $v_l=2$ case.
For the latter the energy $E(m)$, for $G\to\infty$, remains trapped
in between two s.p.l.: hence the $G$-dependent
normalisation constant $C_1(m)$ will tend to a 
finite value $C_1^\infty(m)$. Accordingly
the matrix element (\ref{eq:az13}) 
will vanish with $G$.

On the contrary, in the $v_l=0$ case, where the unperturbed pair lives on the
lowest level, $E(1)\sim -G\Omega$. 
Hence, for large $G$, since $C_1^{\infty}(1)\sim G\sqrt{\Omega}$, one finds
\begin{equation}
  \label{eq:az34}
  <(1)|\hat A^\dagger|0>\sim\sqrt{\Omega}~,
\end{equation}
in accord with \eqref{eq:ax11}.
Thus at large $G$, for this matrix element, $v_l$ and $v$
coalesce.

Next we study the transition from one to two pairs, namely
the matrix element of $\hat A^\dagger$
between ${\cal B}^*(m)$ and ${\cal B}_1^*(m_1,m_2){\cal B}_2^*(m_1,m_2)$.

A lengthy calculation yields
\begin{multline}
  \label{eq:ax5}
  <(m_1,m_2)|\hat A^\dagger|(m)>=\\
  \frac{\sum_{\mu\nu}
  [\tilde \beta^{(1)}_\mu(m_1,m_2)\tilde \beta^{(2)}_\nu(m_1,m_2)]^*
  \left[\tilde \beta_\mu(m)\sqrt{\Omega_\nu}
    +\tilde \beta_\nu(m)\sqrt{\Omega_\mu}
  -\dfrac{2}{\sqrt{\Omega_\mu}}\beta_\mu(m)\delta_{\mu\nu}\right]}
{{\cal N}}~,
\end{multline}
the normalisation constant reading
\begin{multline}
  \label{eq:ax6}
  {\cal N}^2=1+\sum_{\mu\nu}\left[\left(\tilde \beta^{(1)}_\mu(m_1,m_2)
        \tilde \beta^{(2)}_\nu(m_1,m_2)\right)^*\tilde \beta^{(2)}_\mu(m_1,m_2)
      \tilde \beta^{(1)}_\nu(m_1,m_2)\right.\\
      \left.-\dfrac{2}{\Omega_\mu}
      \left|\tilde \beta^{(1)}_\mu(m_1,m_2)\right|^2
      \left|\tilde \beta^{(2)}_\mu(m_1,m_2)\right|^2\delta_{\mu\nu}\right]~.
\end{multline}
It will be computed in Appendix A.

Using the Richardson equations to get rid of the sums, we recast 
\eqref{eq:ax5} as follows
\begin{equation}
  \label{eq:ax51}
   <(m_1,m_2)|\hat A^\dagger|(m)>=
   \frac{1}{{\cal N}}
   \frac{2C_1(m_1,m_2)C_2(m_1,m_2)C_1(m)}{G[(E_1(m_1,m_2)-E(m)]
[(E_2(m_1,m_2)-E(m)]}~,
\end{equation}
which is real since $E_1$ and $E_2$ are either real or complex 
conjugate.

\begin{figure}[htp]
  \begin{center}
        \mbox{
      \begin{tabular}{cc}
        \epsfig{file=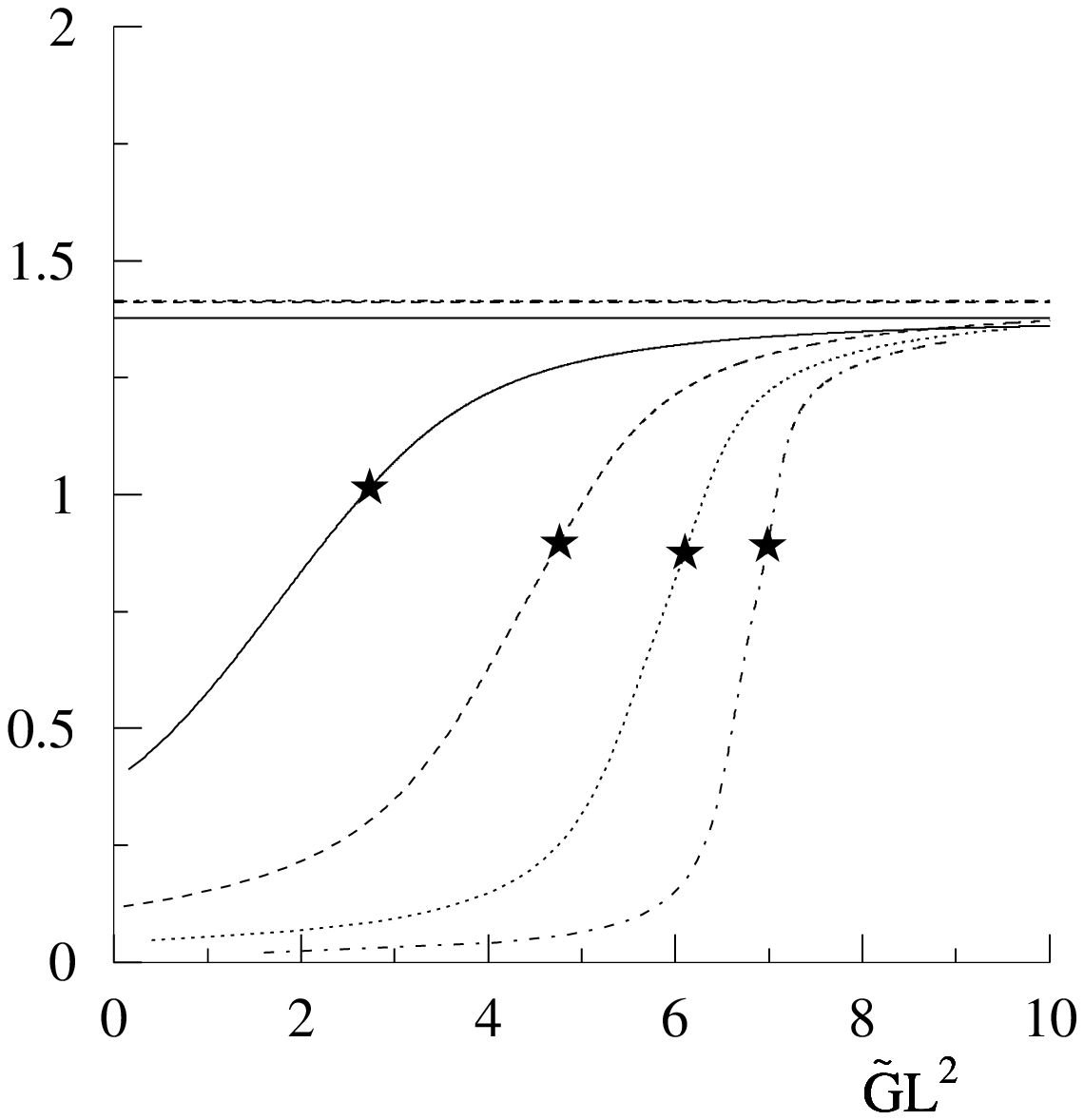,width=6cm,height=7cm}&
        \epsfig{file=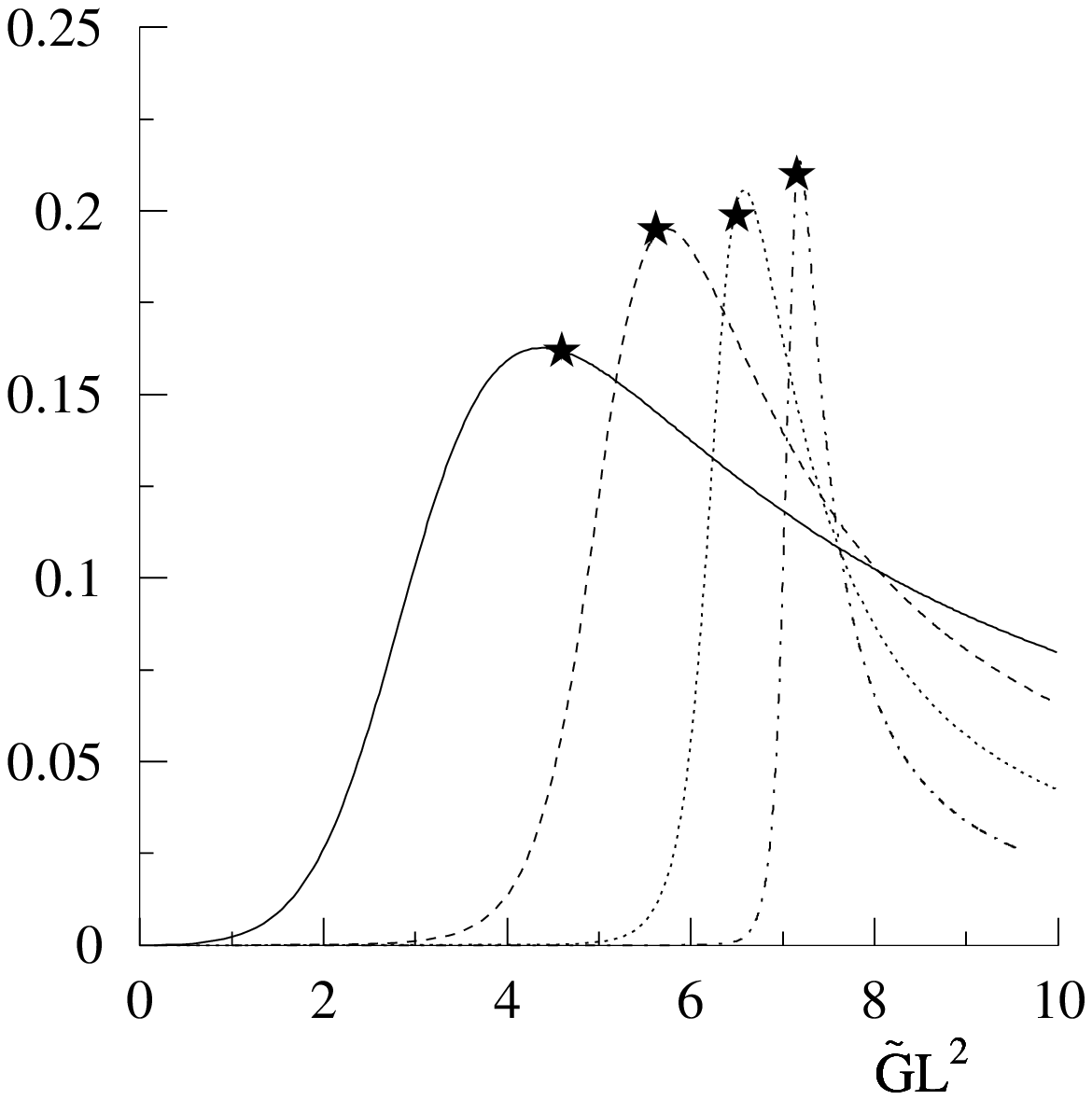,width=6cm,height=7cm}
        \\
        \epsfig{file=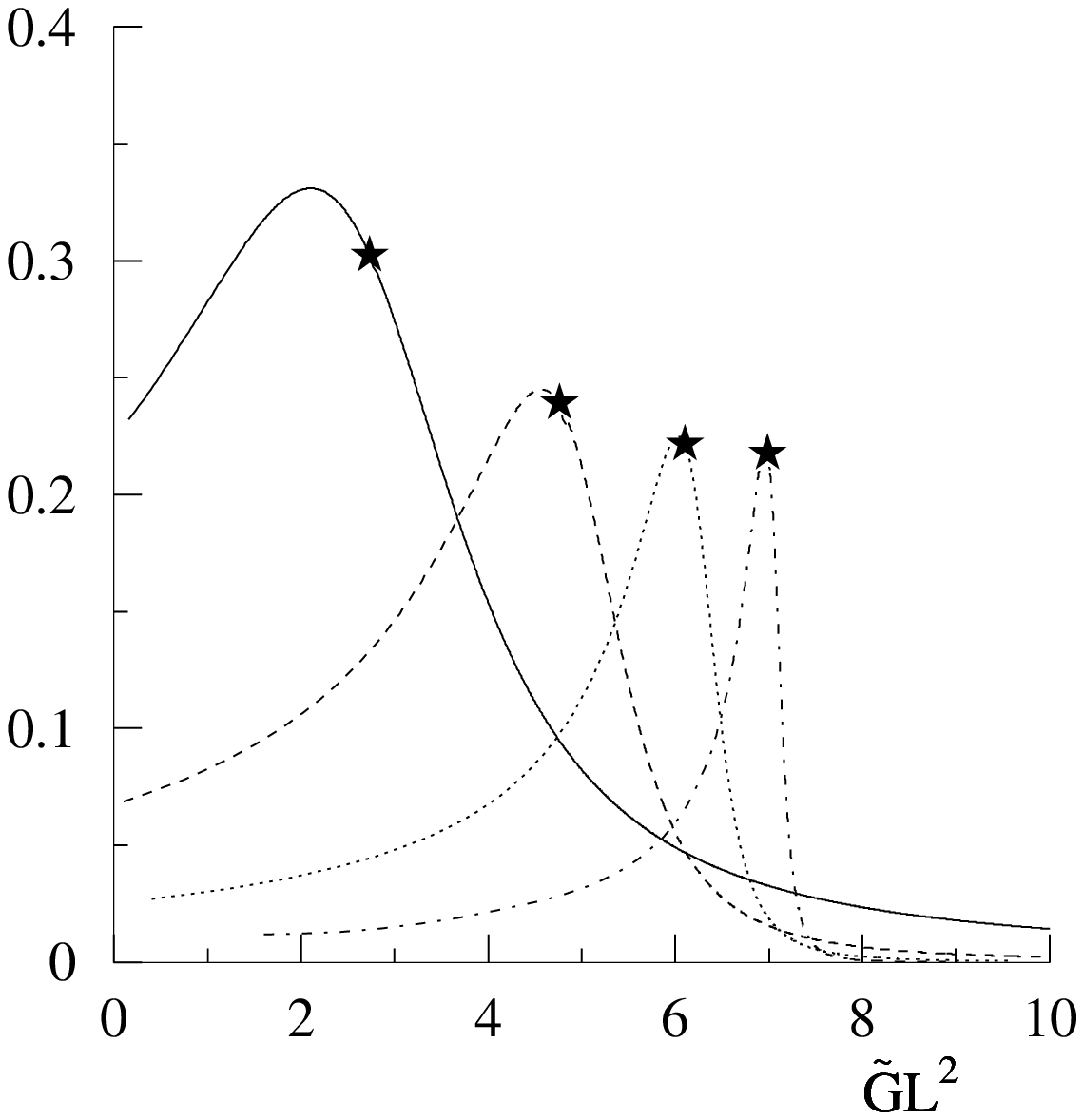,width=6cm,height=7cm}&
        \epsfig{file=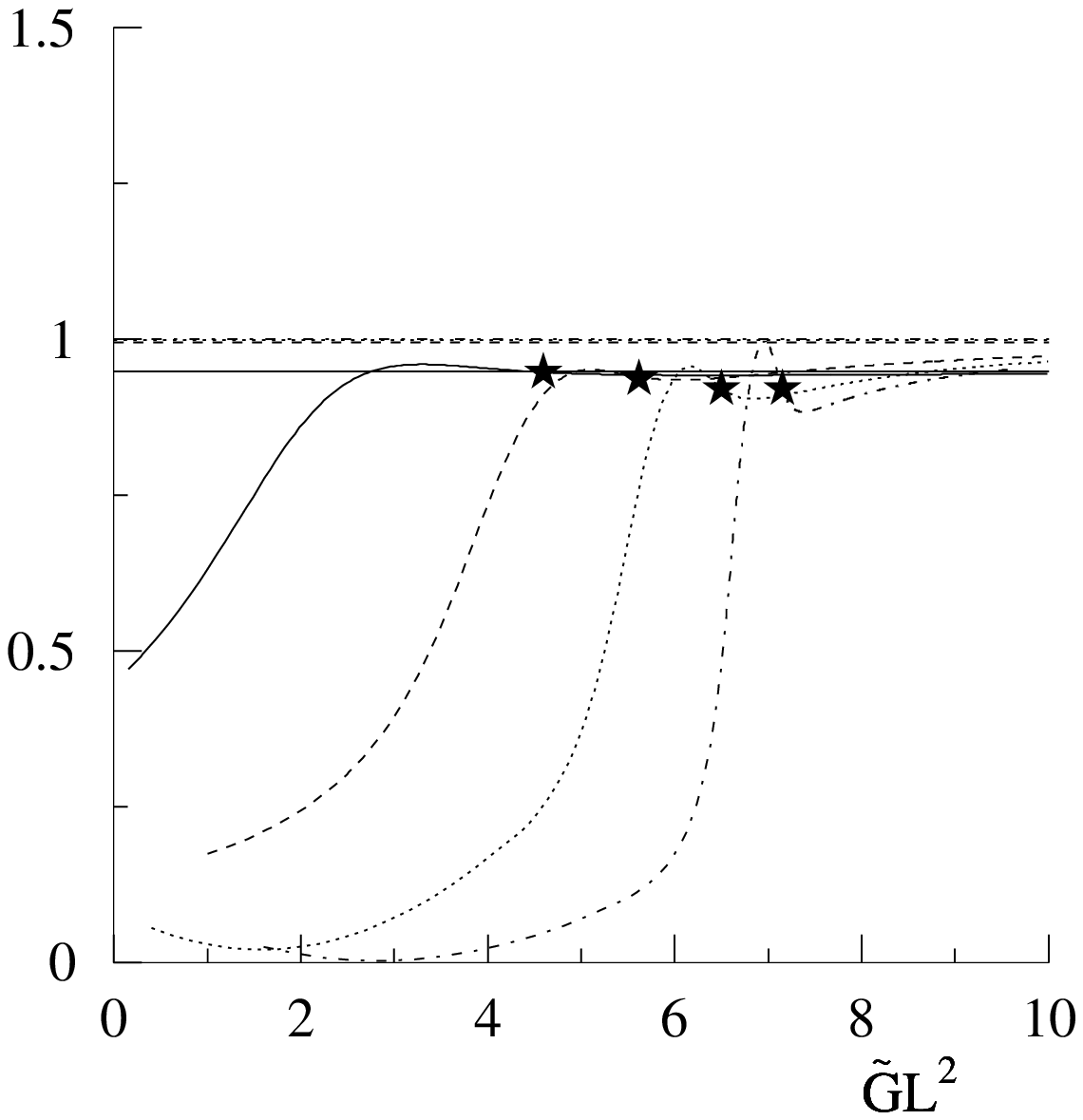,width=6cm,height=7cm}
        \end{tabular}
      }
    \caption{The transition matrix elements \eqref{eq:ax51} divided by 
     $\sqrt{\Omega}$ for $L=$4 (solid), 10 (dashed), 20 (dotted) and 40
      (dot-dashed) levels of a h.o. well.  
      Upper left panel: transition from $m=1$ (collective)
      state to $(m_1,m_2)=(1,2)$ ($v_l=0$) state;
      upper right panel: transition from $m=1$ (collective)
      state to $(m_1,m_2)=(2,2)$ ($v_l=2$) state;
      lower left panel: transition from $m=2$ (trapped)
      state to $(m_1,m_2)=(1,2)$ ($v_l=0$) state;
      lower right panel: transition from $m=2$ (trapped)
      state to $(m_1,m_2)=(2,2)$ ($v_l=2$) state.
      An asterisk denotes the position of the critical points.
      The straight lines represent the asymptotic values for $G\to\infty$.
}
    \label{fig:7}
  \end{center}
\end{figure}

We display in 
fig.~\ref{fig:7} the amplitudes \eqref{eq:ax51} 
for the transitions from a $n=1$ to a $n=2$ system
(divided, for obvious convenience, by $\sqrt{\Omega}$).
We consider  
the $n=1$ system either in the ground collective state 
($m=1$) or in the first excited trapped state ($m=2$).
On the other hand the $n=2$ system is either in the $v_l=0$
(namely $(m_1,m_2)=(1,2)$) or in the $v_l=2$ ($(m_1,m_2)=(2,2)$) state.
The calculation is performed for a harmonic oscillator well with $L=4,10,20$ and $40$.

To understand the behaviour of the curves 
first consider the weak coupling limit, namely (from eq. \eqref{eq:ax5})
\begin{equation}
  \label{eq:ax22}
  <(m_1,m_2)|\hat A^\dagger|(m)>~\xrightarrow[G\to0]{}~
  \frac{\sqrt{\Omega_{m_1}}\delta_{m_2m}+\sqrt{\Omega_{m_2}}\delta_{m_1m}
    -\dfrac{2}{\sqrt{\Omega_{m_1}}}\delta_{m_1m}\delta_{m_2m}}
  {\sqrt{1+\left(1-\dfrac{2}{\Omega_{m_1}}\right)\delta_{m_1m_2}}}~.
\end{equation}
The above yields:
\begin{eqnarray}
\sqrt{3} &\mbox{for the transition}& m=1\,\rightarrow\,(m_1,m_2)=(1,2)\,,
\nonumber\\
0 & \mbox{for the transition}& m=1\,\rightarrow\,(m_1,m_2)=(2,2)\,,
\nonumber\\
1 & \mbox{for the transition}& m=2\,\rightarrow\,(m_1,m_2)=(1,2)\,,
\nonumber\\
2 & \mbox{for the transition}& m=2\,\rightarrow\,(m_1,m_2)=(2,2)\,.
\nonumber
\end{eqnarray}

The behaviour around the critical points 
(which of course depend upon $L$)
is quite delicate since both the numerator and the denominator in 
\eqref{eq:ax5} 
are singular when $E_1$ and $E_2$ tend to $2 e_1$.
In fact the numerator in \eqref{eq:ax5} vanishes
like $x^2$ (see eqs. \eqref{eq:ab36} for the definition of $x$), 
but so does the denominator. 
In the end the transition \eqref{eq:ax5}, for small $x$, is found to read
\begin{equation}
  \label{eq:bx21}
  <(m_1,m_2)|\hat A^\dagger|(m)>~\xrightarrow[G\to G_{\rm cr}^\pm]{}~
  \dfrac{2}{G_{\rm cr}^\pm}
  \frac{\sum_{\mu\not=1}
\dfrac{\sqrt{\Omega_\mu}\tilde 
      \beta_\mu(m)}{2\epsilon_\mu-2\epsilon_1}
\pm\tilde \beta_1(m)
    {\cal P}_{(2)}}
  {\sqrt{6{\cal P}_{(2)}^4\pm 8{\cal P}_{(2)}{\cal P}_{(3)}^3
      +2{\cal P}_{(4)}^4}}~.
\end{equation}
The value of the transition matrix elements at the critical 
value of $G$ is marked by an asterisk in fig. \ref{fig:7}.

As we have seen in Section \ref{sec:Gcr}, the inverse
moments of the level distribution
vanish in the $L\to\infty$ limit: hence \eqref{eq:bx21} diverges.
This occurrence, not appearing in fig.~\ref{fig:7} because of the chosen normalisation,
relates to the ODLRO (off-diagonal long range order) which sets in into a 
system close to a phase transition.

Specifically the diagonal amplitudes for large $L$ behave according to
\begin{eqnarray}
<(m_1,m_2)|\hat A^\dagger|(m)>
\stackrel{L\to\infty} \simeq
  \begin{cases}
\sqrt{2\Omega}\ \theta\left(\tilde G-\tilde G_{\rm cr}\right) \,\,\, v_l=0\\
\sqrt{\Omega}\ \theta\left(\tilde G-\tilde G_{\rm cr}\right) \,\,\,\,\, v_l=2\,.
  \end{cases}
\end{eqnarray}
The off-diagonal amplitudes behave instead as 
\begin{equation}
<(m_1,m_2)|\hat A^\dagger|(m)>
\stackrel{L\to\infty} \simeq
\delta\left(\tilde G-\tilde G_{\rm cr}\right)
\end{equation}

From the figures it is also clear that the critical value of $G$ increases
with $L$ and $\tilde G_{\rm cr} L^2\to 8$ as $L\to\infty$ (see eq.~\eqref{gcrho}).

Finally we consider the large $G$ limit, ruled by
eqs. \eqref{eq:ax51},\eqref{eq:ax53} and \eqref{eq:ax60}.

Here first observe that the normalisation constant ${\cal N}$ 
(see eq.\eqref{eq:ax60} in the Appendix) is always finite,
reading 
\begin{equation}
  \label{eq:ax69}
  {\cal N}\longrightarrow
  \begin{cases}
    \sqrt{2-\frac{2}{\Omega}}&\text{for~}v_l=0\\
    \sqrt{1-\frac{2}{\Omega}}&\text{for~}v_l=2~~.
  \end{cases}
\end{equation}
In deducing the above use has been made of the asymptotic expressions
for $E_1$ and $E_2$.

Considering next the transitions matrix element \eqref{eq:ax51}
from a one pair to a two pairs state with $v_l=0$ or $2$,
we discuss the two cases:
 \begin{description}
 \item[i)]
initial state: $n=1$, $v_l=0$.

For this, at large $G$, the energy behaves as $-G\Omega$ 
and ${\cal N}\sim G\sqrt{\Omega}$.

Now the $n=2$ final state can have:
   \begin{description}
   \item[a)] $v_l=0$. In this case, 
using the expressions for the pair energies 
$E_{1,2}$ and  the normalisation constants $C_{1,2}$, one finds the asymptotic
result $\sqrt{2(\Omega-1)}$, which coincides with the degenerate one
for vanishing $v$ (not $v_l$!);
   \item[b)] $v_l=2$. Here one so\-lu\-tion (say $E_1$) behaves as 
$-G\Omega$, while the other ($E_2$) remains finite, as discussed in 
sec. \ref{sec:vl2}. 
     Specifically, one finds
     $$E_1\simeq -G_{\rm eff}^{(1)}\Omega\simeq -\frac{\Omega^2 G}{\Omega+2}$$
     and $E_2$ much lower than $E_1$ and of the energy of the initial state. 
     The normalisation constants are 
     $C_1=G_{\rm eff}^{(1)}\sqrt{\Omega}$ and $C_2^\infty$ (this one turns
     out to be $G$-independent and $\sim 1$).
     Using the above and the expression \eqref{eq:ax69}
     for ${\cal N}$, one then gets for the matrix element the asymptotic value
     $$\frac{C_2^\infty}{G}\sqrt{\frac{\Omega}{\Omega-2}}~,$$
     which is vanishing as $G^{-1}$ in the strong coupling limit, as it
     should, since it entails a variation $\Delta v_l=2$
     of ``like-seniority''. 

   \end{description}
 \item[ii)] initial state: $n=1$, $v_l=2$.

   For this the asymptotic values of $C$ and of the energy (denoted
   by $C^\infty$ and $E^\infty$) are finite: $C^\infty$ is of order 1
   and $E^\infty$ is of the order of the unperturbed levels.
   Again, the $n=2$ final state may have:
   \begin{description}
   \item[a)] $v_l=0$. In this case
     $E^\infty$ is negligible with respect to $E_1$ and $E_2$, whose
     asymptotic behaviour is known. The matrix element is then found to read
     $$\frac{\sqrt{2}C^\infty}{G\sqrt{\Omega(\Omega-1)}}~.$$
     Again a transition $\Delta v_l=2$ entails a behaviour
     of order $G^{-1}$, but in this case the extra factor $\sim \Omega^{-1}$
     induces a faster decrease of the transition matrix element,
     as clearly apparent in fig. \ref{fig:7} (lower left panel);
   \item[b)] $v_l=2$. Calling $E_2^\infty$
     (and likewise $C_2^\infty$) the $G\to\infty$
     values of the pair energy (and of the normalisation constant)
     of the trapped pair, we find the following
     expressions for the asymptotic matrix element 
     $$
     \frac{2C_2^\infty C^\infty}{\sqrt{\Omega-2}G(E^\infty-E_2^\infty)}~.$$
     Now two different possibilities occur: if $E^\infty$ and  
     $E_2^\infty$ are not both confined in the range $(2e_{\mu-1},2e_\mu)$
     then, in the strong coupling regime, their difference remains finite 
     and the same occurs for the normalisation coefficients 
     $C^\infty$ and $C_2^\infty$. Hence
     the matrix element is ruled by the factor $1/G$ and thus vanishes. 
     Otherwise the difference $E^\infty-E_2^\infty$
     tends to vanish. In fact, let $E(G)$ be the solution of the
     one pair equation \eqref{eq:az20}. Then the trapped pair energy 
     $E_2$, in the strong 
     coupling limit, solves eq. (\ref{eq:49}b), and, as a consequence,
     one has $E_2=E(G^{(2)}_{\rm eff})$, with $G^{(2)}_{\rm eff}$ given by
     eq. \eqref{eq:50b}. Thus, inserting $E_2=E-y$ into eq. (\ref{eq:49}b)
     and expanding in $y$, one finds,  using \eqref{eq:az20},
     $$y=\frac{2|C^\infty|^2}{G\Omega}~.$$
     Hence the $G\to\infty$ limit of the transition matrix element becomes
     $$\frac{C_2^\infty}{C^\infty}\frac{\Omega}{\sqrt{\Omega-2}}~,$$
     where the ratio $C_2^\infty/C^\infty$ can be shown to be given by
     $$\frac{C_2^\infty}{C^\infty}
     =\frac{G}{G^{(2)}_{\rm eff}}
     =1-\frac{2}{\Omega}~.$$
     Therefore the asymptotic matrix element reads
     $\sqrt{\Omega-2}$, namely the value of the degenerate case (\ref{eq:ax11}).
   \end{description}
 \end{description}

\section{Conclusions}

It is well-known that the pairing Hamiltonian, a reduced version of the BCS
model of superconductivity, is solved by the Richardson's equations 
as far as the states of zero seniority $v$ are concerned.

In this paper we have explored the solutions of these equations in the simple
case of only two pairs of fermions, but living in any number of s.p.l. of any
potential well.
In this framework a further classification of the states in terms of 
$v_l$, which accounts for their degree of collectivity, appears convenient.
It is elegant that $v_l$ and $v$ coincide for $G\to\infty$, whereas
at finite $G$ one ($v_l$) classifies the states according to the same 
pattern of the other ($v$) without breaking any pair.
At small $G$, of course, no need for $v_l$ is felt.

Concerning the structure of the solutions we have naturally searched, and
obtained, an expression for the energy of the collective $v_l=v=0$ mode in 
terms of the statistical features of the s.p.l. distribution, namely in
terms of the moments of the latter.
The same we have achieved for the two critical values of the coupling
which mark the transition from the physics of the mean field to the one
of the pairing force.

According to the general theory of Richardson {\em for a two pairs system}
at most two critical values ($G_{\rm cr}^+$ and $G_{\rm cr}^-$) can exist
on an unperturbed s.p.l. provided the pair degeneracy of the latter is one.
Remarkably $G_{\rm cr}^+$ and $G_{\rm cr}^-$ are found to merge in the
$L\to\infty$ limit and here coincide with the $G_{\rm cr}$, previously found
in \cite{BaCeMoQu-02}, for a system of just one pair living on any number
of s.p.l.
Note that in the latter case a $G_{\rm cr}$ should {\em not} exist according
to Richardson, but we have found that in fact it does by following the 
behaviour of the trapped modes with $G$.
Its existence essentially rests on the trace of $\hat H_{pair}$ or,
equivalently, on the balance between the collective and the trapped modes, 
a balance reflecting not only the strength $G$ of the pairing force, but,
critically, the degeneracies of the s.p.l.
In the $n=2$ case the latter markedly affect the critical values of $G$,
which, in turn, signal the onset of the validity of the BCS framework.
Indeed, for example, we have found that the $G_{\rm cr}$ of the $\Omega=1$ 
model is much larger than the one of the harmonic oscillator well.
Whether this will entail the absence in the $\Omega=1$ model of a transition
in the trapped modes for the $n=2$ system, as it happens for the $n=1$
system, is an issue we are currently investigating.

Be as it may, we have explored in detail the transition associated to
$G_{\rm cr}^+$ and $G_{\rm cr}^-$ by following the behaviour with $G$ of
both the excitation energies of the $v_l=2$ and $v_l=4$ states and of the
pair transfer matrix elements, finding for the latter a most conspicuous
enhancement (the more so, the larger $L$) in the proximity of $G_{\rm cr}$,
clearly reflecting the onset of an ODLRO in the system.
Concerning the excitation energies we have found that they closely approach
the BCS predictions at large $G$ especially for the case of two
quasi-particle ($v_l=2$) excitations, notwithstanding that the latter, in the
BCS picture, amounts to break a pair, an occurrence never happening in the
Richardson frame.

Finally we have found that the BCS ground state of the $n=2$ system
in the simple $L=3$ harmonic oscillator model 
has $\Delta/d\simeq 2.5$ in the proximity of the critical values,
being $d$ the spacing among the s.p.l. (actually $\Delta/d$ is growing
linearly with $G$). The Anderson's
criterion~\cite{Anderson} for superconductivity is thus fulfilled.
Indeed in the ground state our system can be viewed as four fermions 
(a ``quartet'') sitting on the same energy level. 

\appendix
\section{Appendix}

In this Appendix we shortly provide analytic expressions for the
normalisation constants, whenever this is possible.
When both the pair energies are real no simple formulas 
for the normalisation coefficients $C_{1,2}(m,n)$ are available,
whereas when $E_1^*=E_2$ one can use eq. \eqref{equazione} to get
\begin{equation}
  \label{eq:ax52}
  C_{1,2}=\left|\frac{E_1-E_2}{2}\right|~.
\end{equation}

The above allows to simplify the expressions for the normalisation constant 
distinguishing, however, whether the solutions are real or complex conjugate. 

In the former case a tedious calculations provides
\begin{equation}
  \label{eq:ax53}
  {\cal N}=\sqrt{1-\frac{2[C^2_1(m,n)+C^2_2(m,n)]}{[E_2(m,n)-E_1(m,n)]^2}}~,
\end{equation}
still in terms of the unspecified normalisation constants $C_1$ and $C_2$.

In the latter case the normalisation is known, but, to get ${\cal N}$,
we need to define
\begin{equation}
  \label{eq:ax55}
   \tilde 
   C_k(m,n)=\frac{1}{\sqrt{\sum_\mu\dfrac{\Omega_\mu}{(2e_\mu-E_k(m,n))^2}}}
   ~,\qquad k=1,2
\end{equation}
where, unlike in \eqref{eq:ax58}, no absolute values appear.
Thus the $\tilde C_k(m,n)$ are complex and
$\tilde C_2^*=\tilde C_1$.
With the help of \eqref{eq:ax55}, and using \eqref{eq:ax52},
one then gets
\begin{equation}
  \label{eq:ax60}
  {\cal N}=\frac{|E_1-E_2|^2}{4|\tilde C_1\tilde C_2|}
  \sqrt{1-\frac{2 (\tilde C_1^2+ \tilde C_2^2)}{(E_1-E_2)^2}}~.
\end{equation}


\newpage
\begin{thebibliography}{99}
\bibitem{BaCeMoQu-02}
M.~B.~Barbaro, R.~Cenni, A.~Molinari and M.~R.~Quaglia, Phys.\ Rev.\ C {\bf 66} 
(2002) 034310.
\bibitem{Sierra:1999rc}
G.~Sierra, J.~Dukelsky, G.~G.~Dussel, J.~von Delft and F.~Braun,
Phys.\ Rev.\ B {\bf 61} (2000) 11890.
\bibitem{Dukelsky:2001fe}
J.~Dukelsky, C.~Esebbag and P.~Schuck,
Phys.\ Rev.\ Lett.\  {\bf 87} (2001) 066403.
\bibitem{Gau95}
M.~Gaudin, ``Modeles Exactement Resolus'', les Editions de Physique, 
France, 1995
\bibitem{Roman}
J.~M.~Roman, G.~Sierra and J.~Dukelsky,
Phys.\ Rev.\ B {\bf 67} (2003) 064510.
\bibitem{Ric-65}
R.~W.~Richardson, Jour.\ Math.\ Phys.\ {\bf 6} (1965) 1034.
\bibitem{Ric-64}
R.~W.~Richardson and N.~Sherman,\ Nucl.\ Phys. {\bf 52} (1964) 221.
\bibitem{Yuz03}
E.~A.~Yuzbashyan, A.~A.~Baytin and B.~L.~Altshuler,
Phys.\ Rev.\ B {\bf 68} (2003) 214509.
\bibitem{Hase}
M.~Hasegawa and S.~Takaki,
Phys.\ Rev.\ C {\bf 35} (1987) 1508.
\bibitem{Roman:2002dh}
J.~M.~Roman, G.~Sierra and J.~Dukelsky,
Nucl.\ Phys.\ B {\bf 64} (2002) 483.
\bibitem{Sch01}
M.~Schechter, Y.~Imry, Y.~Levinson and J.~von Delft,
Phys.\ Rev.\ B {\bf 63} (2001) 214518.
\bibitem{Barbaro:1999fh}
M.~B.~Barbaro, A.~Molinari, F.~Palumbo and M.~R.~Quaglia,
Phys.\ Lett.\ B {\bf 476} (2000) 477.
\bibitem{Bes-Broglia}
D.~Bes and R.~A.~Broglia, Lezioni di Varenna
International School of Physics ``Enrico Fermi'', Course LXIX,
edited by A. Bohr and R.A. Broglia, Varenna, 1977, p.59.
\bibitem{Anderson}
P.~W.~Anderson, J.\ Phys.\ Chem.\ Solids {\bf 11} (1959) 28.
\end{thebibliography}
\end{document}